# Securing AI-based Healthcare Systems using Blockchain Technology: A State-of-the-Art Systematic Literature Review and Future Research Directions

Rucha Shinde[1] · Shruti Patil[2] · Ketan Kotecha[2] · Vidyasagar Potdar[3] · Ganeshsree Selvachandran[4] · Ajith Abraham[5]

## Abstract

Healthcare systems are increasingly incorporating Artificial Intelligence into their systems, but it is not a solution for all difficulties. AI's extraordinary potential is being held back by challenges such as a lack of medical datasets for training AI models, adversarial attacks, and a lack of trust due to its black box working style. We explored how blockchain technology can improve the reliability and trustworthiness of AI-based healthcare. This paper has conducted a Systematic Literature Review to explore the state-of-the-art research studies conducted in healthcare applications developed with different AI techniques and Blockchain Technology. This systematic literature review proceeds with three different paths as natural language processing-based healthcare systems, computer vision-based healthcare systems and acoustic AI-based healthcare systems. We found that 1) Defence techniques for adversarial attacks on AI are available for specific kind of attacks and even adversarial training is AI based technique which in further prone to different attacks. 2) Blockchain can address security and privacy issues in healthcare fraternity. 3) Medical data verification and user provenance can be enabled with Blockchain. 4) Blockchain can protect distributed learning on heterogeneous medical data. 5) The issues like single point of failure, non-transparency in healthcare systems can be resolved with Blockchain. Nevertheless, it has been identified that research is at the initial stage. As a result, we have synthesized a conceptual framework using Blockchain Technology for AI-based healthcare applications that considers the needs of each NLP, Computer Vision, and Acoustic AI application. A global solution for all sort of adversarial attacks on AI based healthcare. However, this technique has significant limits and challenges that need to be addressed in future studies.

1 Symbiosis Institute of Technology, Symbiosis International (Deemed University), Pune, Maharashtra, India.
2 Symbiosis Centre for Applied Artificial Intelligence (SCAAI), Symbiosis Institute of Technology, Symbiosis International (Deemed University), Pune, Maharashtra, India.
3 Curtin University, Perth, Australia.
4 UCSI University, Kuala Lumpur, Malaysia.
5 Machine Intelligence Research Laboratories, Auburn, WA, USA.

## I. Introduction

From infectious diseases, cancer to radiology, the healthcare sector is in a desperate need of transformation. There are almost countless ways towards using technologies to provide more accurate, reliable, and effective treatments. These treatments can be precise at the right time in a clinical decision. Artificial intelligence uses a computer program with precise commands to execute functions that usually require a human intelligence. Algorithms are coded programming rules. Machine learning is a method of constantly improving an algorithm. The improvement process utilizes vast volumes of data and is performed dynamically, enabling the algorithm to adjust to improve the accuracy of said artificial intelligence. AI can understand and interpret language, identify objects, detect sounds, and learn patterns to execute problem-solving operations.

In this review, as shown in Figure 1 we provide insight into three perspectives of Artificial Intelligence and their specific challenges in healthcare: Natural Language Processing (NLP), Computer Vision, and Acoustic AI. Natural language processing's primary objective is for computers to comprehend texts and languages the same way as humans do. Computer systems will interpret, deduce, summarize, translate, and synthesize exact text and language once accomplished. As mentioned in Figure 1’s text section, in healthcare systems, a vast amount of textual data is generated in the form of clinical reports, lab reports, handwritten notes, and other documents like admission, and discharge notes, and many more. To handle this enormous data and analyze it manually would be overweighing tasks for clinical experts. The primary tasks that can be driven through NLP are extracting important facts from text, classification of information, and opinion mining. NLP helps by analyzing these growing data and convert them to a manageable computer format. It can also help to assist clinical decisions, identify critical patients, and classify diseases and disorders.

Nowadays, computer vision is also being used in the healthcare industry. The rapidly growing area of computer vision is concerned with training computers to mimic human vision and understand the items in front of them. Computer vision does this by leveraging artificial intelligence algorithms which analyze images. As shown in Figure 1’s image section, X-ray, CT, MRI, Ultrasound images and videos proved to be the most vital elements in a patient’s diagnosis. Through different tasks like object detection, classification, localization, and analysis from images or videos, computer vision can promote remote patient monitoring, automated diagnosis, and automated lab reports. It can promote the emergence of numerous applications that can be lifesaving for patients in Radiology, Oncology, Cardiology, Dermatology, and Fundoscopy.

Along with text and images as illustrated in Figure 1’s audio section, certain sounds like coughing, breathing, heartbeats, crying, etc., play a major role in diagnosing respiratory diseases, pulmonary diseases, cardiac diseases, and pain in neonates. AI assists in automating these diagnoses by detecting sounds, performing classification, and analyzing them through the audio spectrum. Various state-of-the-art deep learning algorithms are available for audio signal processing, which can be helpful in the healthcare industry. It assists in automatic diagnosis of respiratory and pulmonary diseases, diagnosis of cardiac diseases, measuring pain in neonates and detecting depression in human beings.

There are certain common general challenges faced by AI models as mentioned in Figure 1 for their wider adoption in the healthcare industry. First, AI models can work precisely when trained on sufficiently large datasets. Hence the availability of vast, accurate, and trusted datasets for training is one of the major challenges. Second, it can be possible if we decide to aggregate data from different resources. As organizations continue to collect, store, and transport individuals' health sensitive data, it should be protected from privacy violation and security breaches. Third, AI models are black box in nature; hence, it is difficult to identify biased models. There must be the provenance of prediction or classification resulting in specific healthcare input to overcome lack of trust on learned model. Human lives are at stake if the wrong treatment is followed based on the AI results. Fourth, to overcome the threat of rogue devices, there should be secure resource sharing. Fifth, knowledge sharing among researchers and clinical expertise has information privacy issues. Hence, there must be a proven strategy to overcome these challenges to see the healthcare industry dominated by AI in the coming years.

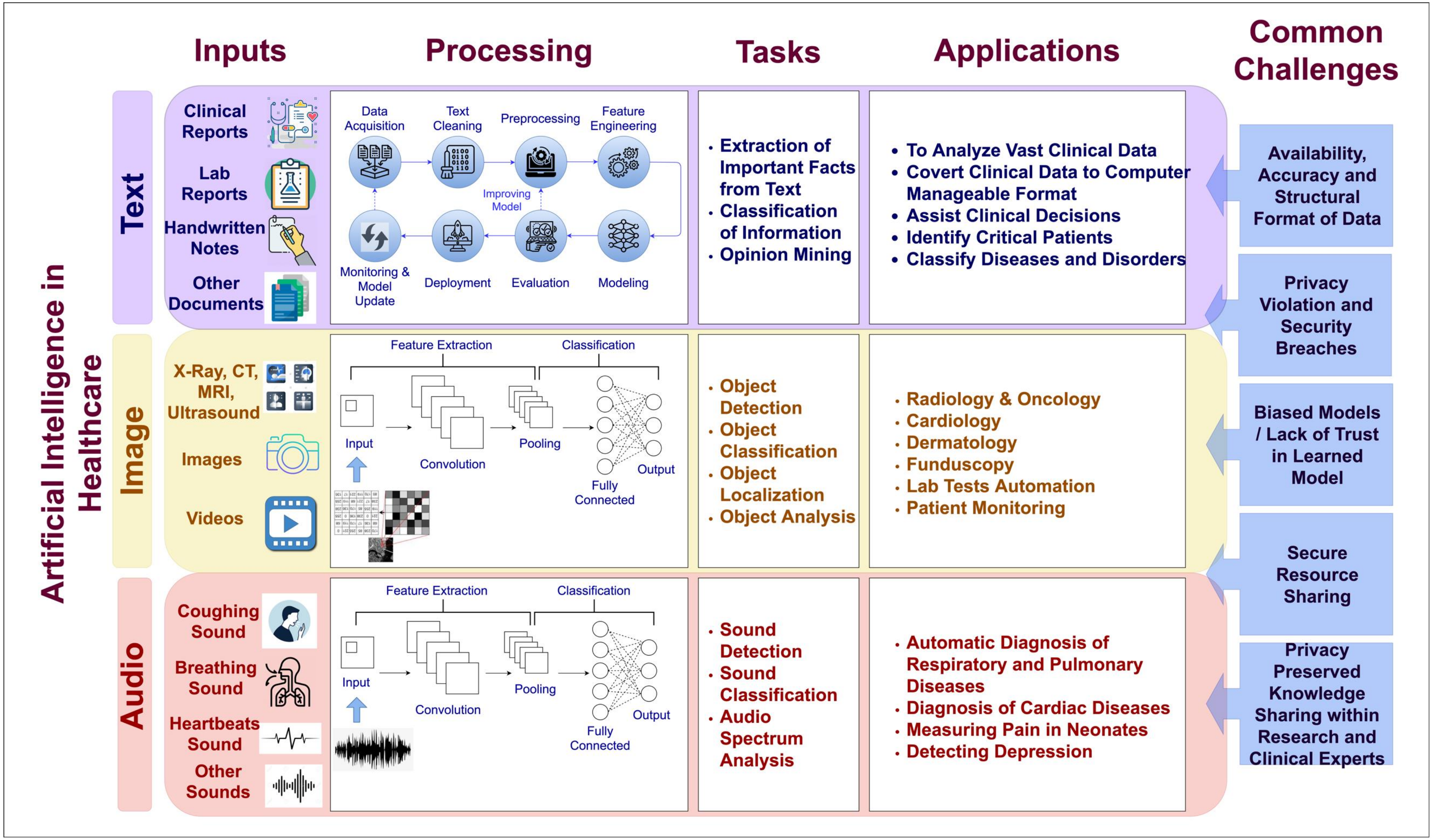

**Figure 1**. Artificial Intelligence in Healthcare.

### A. Background/Rationale

Blockchain technology can address the challenges faced by AI as mentioned in Figure 1 in several ways. A blockchain is a distributed ledger with transactions that are replicated throughout the blockchain's ecosystem. Security and privacy feature of blockchain are enriched with the cryptographic linkage of information in chronicle order, consensus protocol within the network, and smart contracts. Moreover, it builds strong trust among the users. Hence it can also establish trust, organize data, allow sharing resources in AI-based healthcare. This study mainly targets the applicability of blockchain in AI-based healthcare systems considering security and privacy issues in three vertical aspects of it as natural language processing, computer vision and acoustic AI.

Blockchain technology is featured with distributed ledger in the peer-to-peer network. Distributed ledger maintains transactional records securely. This feature promotes the secure distributed learning or federated learning on heterogeneous data by recording local gradients on blockchain. Moreover, a smart contract automates the transaction execution in the distributed network without any third-party or centralized authority. Smart contract is an executable code available at every node which gets triggered on transaction initialization. Smart contracts validate the transaction. Through smart contracts access control rules can be imposed for data access. User provenance is possible with smart contract. For transactional data, a block is generated. Miners are responsible for committing the block in the blockchain using a consensus algorithm. Consensus algorithms mine the block. It makes miners solve difficult cryptographic puzzles and share their results with a group of miners. The miner who first solves the puzzle gets a chance to mine a block of the transactions into the existing chain of blocks and replicate the new chain at every node. For collective decision making on diagnosis and treatment in AI-based healthcare systems consensus algorithms can be the proven technique. Blocks are linked with each other cryptographically, which makes them immutable and auditable. The same copy of the ledger is replicated at all nodes in the network, henceforth achieving the highest degree of availability and transparency. Cryptographic linkage can validate the medical data and support tamperproof copy of it. There are three types of blockchain available as public blockchain, private blockchain, and consortium blockchain. In public blockchain, anyone can enter the network and participate in the transaction process. In contrast, private blockchain restricts entry without proper authentication and verification. Consortium blockchain combines the features of public and private blockchain. Figure 2 depicts the flow of activities in the blockchain network.

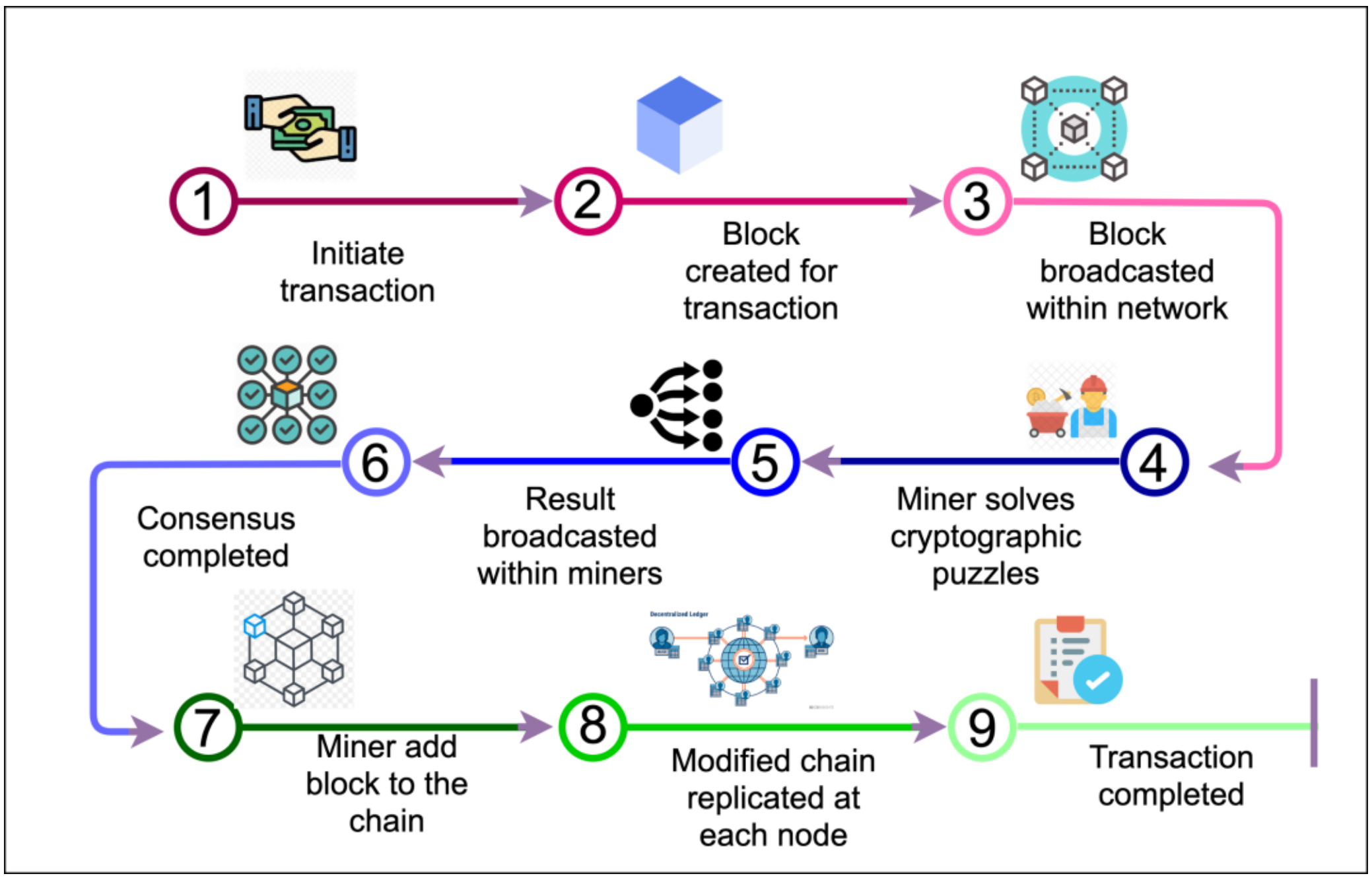

**Figure 2.** Blockchain Process.

### B. Significance of Study:

In AI-based health care, data-driven learning and exploration will improve awareness and productivity in terms of precise diagnosis and treatment ahead. While giving rise to certain challenges like anonymity, data control, and the revenue generation from patient's sensitive information. It would be critical to establish a strong trust in the data used to boost machine learning tools. To ensure that the resulting treatment quality is high and that patients receive the assistance they need to treat chronic diseases or acute illnesses in a timely manner. Artificial intelligence and machine learning in medicine have enormous potential to improve healthcare, but different adversarial attacks on NLP, computer vision, and acoustic AI are possible which limit their real time adoption. These attacks cannot be tolerated in sensitive application areas like in healthcare.

Blockchain will protect against adversarial attacks considering the security requirements of NLP, computer vision, and acoustic AI, respectively. When it comes to security and privacy, the convergence of blockchain and AI-based healthcare has the potential to be transformative. This study uncovers existing blockchain integrated AI-based healthcare applications. This study maps the blockchain solution to the adversarial attacks in NLP, computer vision, and acoustic AI to bring robustness to AI-based healthcare systems. A conceptual framework is synthesized for the same. This contribution will help to further academic knowledge in the blockchain for AI envisioned healthcare applications.

### C. Terms and Terminology:

- Blockchain: A blockchain is a growing collection of data known as blocks that are connected via encryption. Blockchains are generally administered via a peer-to-peer network for a publicly distributed ledger, with nodes communicating and validating new blocks using a protocol.
- Encryption: Encryption is the process of translating data or information into the code, particularly to restrict data leakage.
- Nodes: The computers in a blockchain network are known as blockchain nodes, and they are responsible to maintain a replica of the distributed ledger and operating as communication channels for different network activities.
- Block: Blocks are used to store batches of valid transactions that have been hashed. Each block contains the cryptographic hash of the previous block on the blockchain. Next, a chain is formed by the connected blocks. This recursive procedure validates the preceding block's integrity to the first block, the genesis block.
- Transaction: An exchange or transfer of assets that occurs within two or more individuals and generates a contractual relationship is referred to as a transaction.
- Cryptographic hash: Cryptographic hash is fixed size of value generated from arbitrary size of data by applying cryptographic functions.
- Genesis block: The Genesis Block can be called as Block 0 as it is the very first block in a blockchain, on which the subsequent blocks are built.
- Smart Contract: A smart contract is a computer program, or a transaction protocol designed to execute, control, or document contractually important events and acts under the conditions of a contractual agreement.
- Consensus Algorithm: A consensus algorithm is a computer program that allows distributed processes or systems to agree on a single data value.
- Natural Language Processing: Natural language processing (NLP) is a field of artificial intelligence (AI) focused on providing computers with the potential to interpret text and spoken languages in much the same way as humans can.
- Computer Vision: Computer vision is an artificial intelligence area that teaches computers to interpret and comprehend the visual environment. Machines can effectively detect and categorize items using digital pictures from cameras and movies and deep learning models — and then respond to what computers perceive.
- Acoustic AI: Acoustic AI uses artificial intelligence to build sophisticated audio signal processing algorithms that understand sounds in the environment on an immense scale.
- Adversarial Attack: An adversarial attack is a strategy for locating a perturbation that alters a machine learning model's prediction. The disturbance may be extremely tiny and invisible to human sight.
- Federated Learning: Federated learning is a machine learning approach in which an algorithm is trained across several decentralized edge devices or servers, keeping local data samples without being exchanged.
- Explainable AI: Explainable AI is a collection of tools and frameworks designed to assist in understanding and interpreting predictions generated by machine learning models.

### D. Evolution of Blockchain in Healthcare 4.0 and Start of Healthcare 5.0:

Healthcare 4.0 has underlined global healthcare with real-time monitoring and involves AI and data analytics. In 2016, blockchain was suggested first time in the healthcare system to liberate inefficient assets, address critical organizational issues, and manage electronic transfers and the exchange of healthcare records. Drug manufacturing systems can also be monitored[1]. MedRec framework based on blockchain technologies was proposed to manage EMRs (Electronic Medical Records)[2]. Blockchain provided trusted data marketplaces [3]. Later considering the privacy issues in public networks, the privacy-preserved healthcare data management system MediBchain was proposed [4]. From 2018, the new approach of integrating Blockchain with AI and IoT (Internet of Things) brought a revolutionary change in healthcare system. This integration has offered more efficient EHR (Electronic Health Record) management, remote patient monitoring, autonomous diagnosis, distributed parallel computing for precise medicine [5]–[9]. In 2019, A prototype of a telemedicine framework using blockchain was developed for securing the remote delivery of healthcare services to rural and remote area in Bangladesh and protecting patient's health related sensitive data[10]. A revolutionary blockchain-based fog monitoring framework was proposed in 2019 for human activity identification. It is an extension to e-Healthcare facilities, based on creating clustered based feature vectors. Remote patient monitoring system is featured with big data analytics [11], [12]. Later in 2020, the convergence of blockchain has extended to federated learning, explainable AI, 5G and 6G network [13], [14]. Protection and privacy are important during the collection, management, and distribution of EHR data [15]. Synthesis of AI and blockchain provide a blueprint for a blockchain-assisted open bionetwork of private healthcare records to accelerate emerging methodologies for medication development and preventative healthcare [16]. The start of the Healthcare 5.0 era has been announced with the Intelligent Telesurgery technology with 6G-enabled Tactile Internet (TI) built on the blockchain to provide real-time and intelligent ultra-responsive healthcare facilities virtually with high effectiveness and productivity [17]. In early 2021, research in secured image processing and sharing is uplifted with blockchain[18]. For the IoMT (Internet of Medical Things) environment, deep learning (DL) with blockchain-assisted secured image transmission and diagnostic model is invented [19]. Again, to provide adequate protection against all cyber-attacks in the healthcare industry, the Hyperledger Fabric technology has proved effective [20]. Figure 3 highlights the milestones in Healthcare 4.0 and the start of Healthcare 5.0 with blockchain technology.

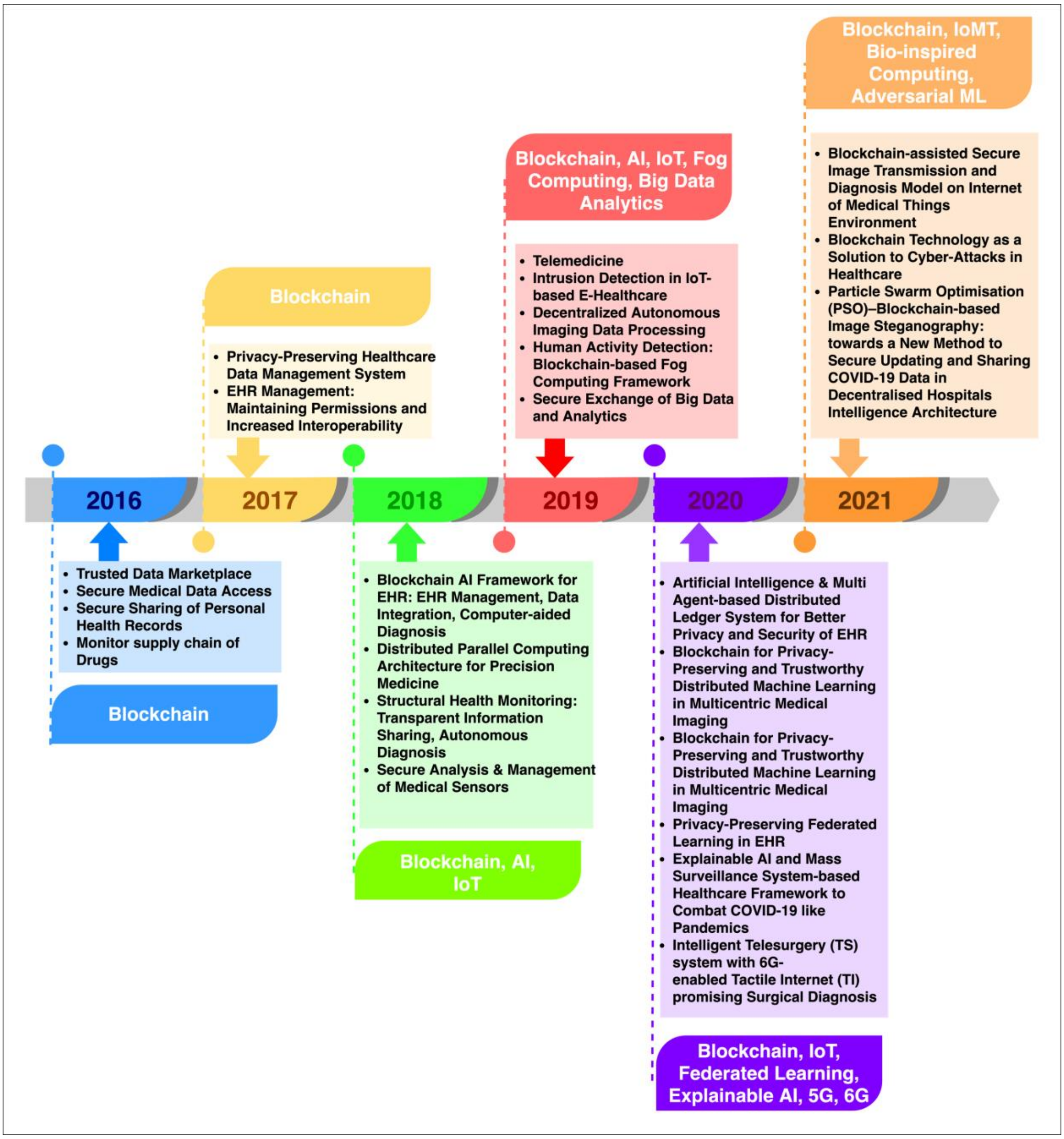

**Figure 3.** Blockchain Evolution in Healthcare 4.0.

## II. Prior Research

There are some relevant literature reviews in the Blockchain and Healthcare space. The review [21] looked at the handling of patient data and identity. The focus was on Electronic Health Records (EHR) and Patient Health Records (PHR) and how blockchain grants patients' power for their data and self-sovereign identity. The study [22] highlighted the blockchain-enabled healthcare applications. In that review, blockchain adaptability in healthcare was also investigated and validated. The study [23] focus on Smart contracts for EHR access control in healthcare considering permissioned and permissionless Blockchain. The reviews [24]–[26] have investigated various blockchain implementations in the healthcare domain and proposed potential research directions and trends in healthcare as precise diagnosis, cybercrime protection and enhance patient care in case of emergencies and remote patient monitoring. Using blockchain technologies in the healthcare sector could enhance information security by allowing healthcare data to be processed and shared while maintaining data privacy and security was concluded by the study [27]. After a thorough examination of the existing major issues in the healthcare field, [28] this review examined the capacity of blockchain technologies to improve the protection, anonymity, and interoperability of healthcare data. This study suggested few innovative blockchain applications in the health care sector, such as blockchain collaboration, smart contract-based health care intelligent claim processing, authorization, and wearable fitness device integration and health tracking. The studies [29], [30] identified the effectiveness of blockchain for healthcare and opportunities and challenges, emphasizing Telemedicine, Telehealth, and E-Health. The scope of the review [31] is restricted to information on blockchain for clinical trials and challenges. [32] Focused on blockchain applications exclusively for EHR. Table 1 depicts the comparison of existing reviews on blockchain in healthcare.

**Table 1.** Comparison of existing literature reviews on blockchain in healthcare.

| Article | Focus on artificial intelligence-based healthcare application | Uncovered cyber-attacks/ Adversarial attacks | Specified type of blockchain platform | Mentioned issues of integrating blockchain into healthcare | Provided solution for issues in integration of blockchain with healthcare system |
|---|---|---|---|---|---|
| [22] | N | Y | Y | Y | Y |
| [25] | N | N | N | Y | N |
| [26] | N | N | Y | Y | N |
| [27] | N | N | Y | Y | Y |
| [28] | N | N | N | Y | N |
| [32] | N | Y | Y | Y | Y |
| [24] | N | N | Y | Y | N |
| [21] | N | N | N | Y | N |
| [23] | N | N | Y | Y | Y |
| [29] | N | N | N | Y | N |
| [31] | N | N | Y | Y | N |
| [30] | N | N | N | Y | N |
| **Our study** | Y | Y | Y | Y | Y |

We note few shortcomings in the previous studies, which can be summarized as follows:

- Previous studies mainly focused on EHR and some specific specialized healthcare services, e.g., Telemedicine.
- Existing literature is not concentrated on blockchain for AI-enabled healthcare. Henceforth, it does not go into depth about blockchain to mitigate adversarial attacks or cyber threats in AI-enabled healthcare.

- Furthermore, none of the surveys mentioned using blockchain technology to incorporate robustness in Natural Language Processing, Computer Vision, and Acoustic AI domains for automated and precise healthcare services.

It is not the first review in blockchain for healthcare, but it is significantly distinct from existing surveys. Our SLR is comprehensive in highlighting innovation, strategies, and threats related to state-of-the-art blockchain-envisioned artificial intelligent healthcare applications targeting Natural Language Processing, Computer Vision, and Acoustic AI domains by considering the potential of AI in healthcare along with the adversarial attacks they may face. In addition, our SLR emphasizes the research gaps to highlight prospective research pathways.

## A. Motivation:

Human intelligence and their physical abilities are restricted with certain boundaries, which leads to automation in healthcare. The Healthcare sector is receiving overwhelming breakthroughs in digital transformation with artificial intelligence. Researchers are coming up with innovative AI solutions in healthcare system to provide better clinical diagnosis and treatment. As shown in Figure 4, over the last five years, research articles in AI-based healthcare are rapidly increasing in the Scopus database (Limited to English articles only). In 2021, the World Health Organization (WHO) has issued ethics and governance standards for implementing artificial intelligence in healthcare [135]. This initiative aims to confirm that these innovations are in line with the objectives of fostering inclusive and fair global health, adhering to health and safety standards, and assisting to achieve sustainable development in Healthcare. The overall investment towards healthcare AI by the public and private sectors is astounding. Accenture estimates that by 2026, the edge AI applications will save $150 billion annually [136].

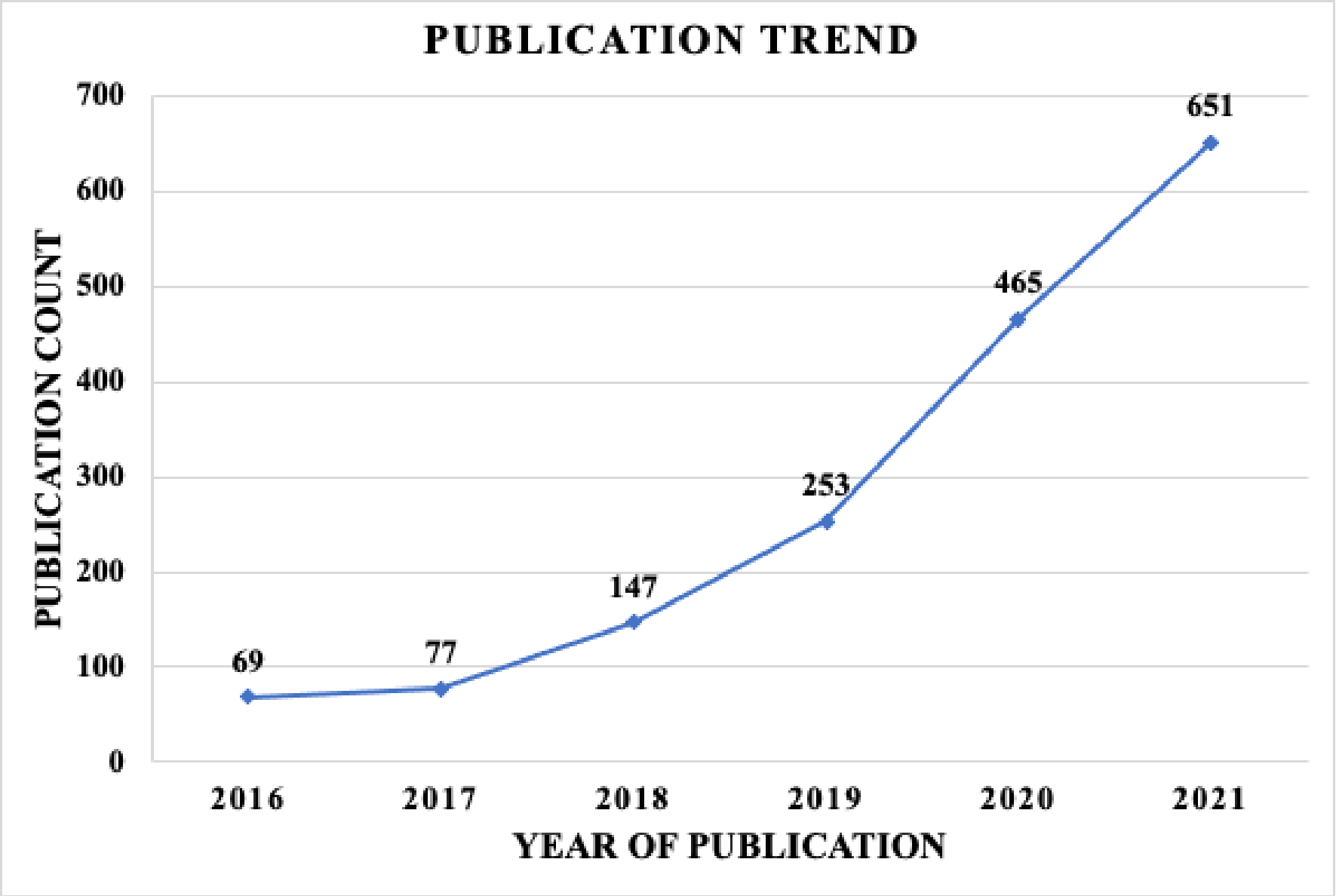

**Figure 4.** Year-wise publication trend for AI in Healthcare (2016-2021) Source: www.scopus.com (Accessed 15/11/2021).

However, it is unclear how often patients will trust AI resources and be inclined to obey an AI diagnosis or adopt the treatment by AI. Users must trust that the algorithm is based on sound clinical guidelines. The data that strengthens these tools is precise, relevant, transparent, and reliable for AI to survive in healthcare. Although artificial intelligence tools are inherently complex, healthcare developers must provide trust and transparency to the fullest in diagnosis and treatment. This study targets blockchain technology as a solution to the shortcomings mentioned earlier. In a pandemic, a large medical workforce is required; however, Intelligent AI-based healthcare automation will alleviate the strain on the healthcare workforce. The adoption of AI-based healthcare may indeed be aided by blockchain technology. However, the work done in blockchain to improve healthcare AI processes is not comprehensive. This study details the advances in AI-based healthcare made possible by blockchain technology and potential directions that will encourage researchers to investigate blockchain technology for AI-based healthcare systems with a new perspective of breaking down the domain into three as NLP, computer vision and acoustic AI.

## B. Research Goal:

This study highlights privacy and security aspects of artificial intelligence-based healthcare systems and risk mitigation with blockchain technology. For this study, healthcare applications from three domains like NLP-based healthcare, computer vision (CV)-based healthcare, and acoustic AI-based healthcare are considered. To achieve the objectives of this systematic literature review, we have formulated four research questions as shown in Table 2.

**Table 2.** Research Goals.

| Sr. No. | Research Question | Objectives | Answered in Section |
|---|---|---|---|
| 1 | How can AI improve the traditional healthcare system? | The objective is to investigate how Natural Language Processing, Computer Vision, and Acoustic AI have impacted healthcare. | VI. B<br>VII. B<br>VIII.B |
| 2 | What are the potential vulnerabilities/threats in AI which make it difficult to adapt AI-based healthcare applications in real-time? | The goal is to examine various potential attacks on Artificial Intelligence that may limit its adoption. | V<br>VI.C<br>VII.C<br>VIII.C |
| 3 | How can blockchain help to enhance AI-based healthcare applications? | The aim is to investigate how blockchain can improve the robustness of AI-based healthcare as blockchain technology could address privacy and security issues in AI. | IV<br>IX |
| 4 | What are the advancements in blockchain technology that help AI healthcare? | Another objective is to explore advancements in Blockchain that can be compatible with AI-based Healthcare. | X.C |

### C. Contribution of the Work:

In this work, we concentrated on NLP, computer vision, and acoustic AI-based healthcare. We have listed various healthcare applications in an individual domain like NLP-based Healthcare, computer vision-based healthcare, acoustic AI-based healthcare, and their challenges. We have touched on adversarial attacks on NLP, computer vision, and acoustic AI, which threaten the use of AI in healthcare. To address security and privacy concerns in AI, authors have reviewed existing blockchain research for AI-based healthcare. Furthermore, we have offered different blockchain algorithms and techniques for mitigating adversarial attacks on NLP, computer vision, and acoustic AI. As blockchain technology has the potential to deal with security and privacy issues, it can play a major role in making more robust AI-based healthcare systems. Finally, we discussed the challenges and constraints of adopting blockchain in healthcare, as well as future research direction in Blockchain AI- envisioned Healthcare.

### D. Paper Organization

Our manuscript is laid out into blockchain for improving the robustness of 3 main domains like natural language processing, computer vision, and acoustic AI, especially for healthcare applications, as shown in the Figure 5. Section I has covered the significance of the study, terms and terminologies, and the evolution of blockchain in healthcare 4.0 and the start of healthcare 5.0. In section II, we have provided a comparison of existing surveys for blockchain in healthcare, also talked about motivation, research goals, the contribution of the work, and paper organization. Section III describes research methodology based on selection criteria, quality assessment, and selection results. In Section IV, we have provided a literature review of blockchain for AI-based healthcare considering challenges. Section V describes the attack surface of artificial intelligence as data, classifier/algorithm, and model. Section VI briefly describes natural language processing, including its techniques and applications in healthcare, adversarial attacks on NLP, and defense. Section VII provides detailed information about computer vision with its techniques and applications in healthcare, adversarial attacks on computer vision, and defense. Section VIII depicts acoustic AI with its techniques and healthcare applications, adversarial attacks on acoustic AI, and work done so far in defense. Then in section IX, we have proposed a blockchain solution for AI-based healthcare considering NLP-based healthcare, computer vision-based healthcare and acoustic AI-based healthcare. Section X discusses the survey outcome, challenges in healthcare industry to adopt blockchain technology in India, and finally, advancements in blockchain. At last conclusion of this systematic literature review is mentioned in Section XI.

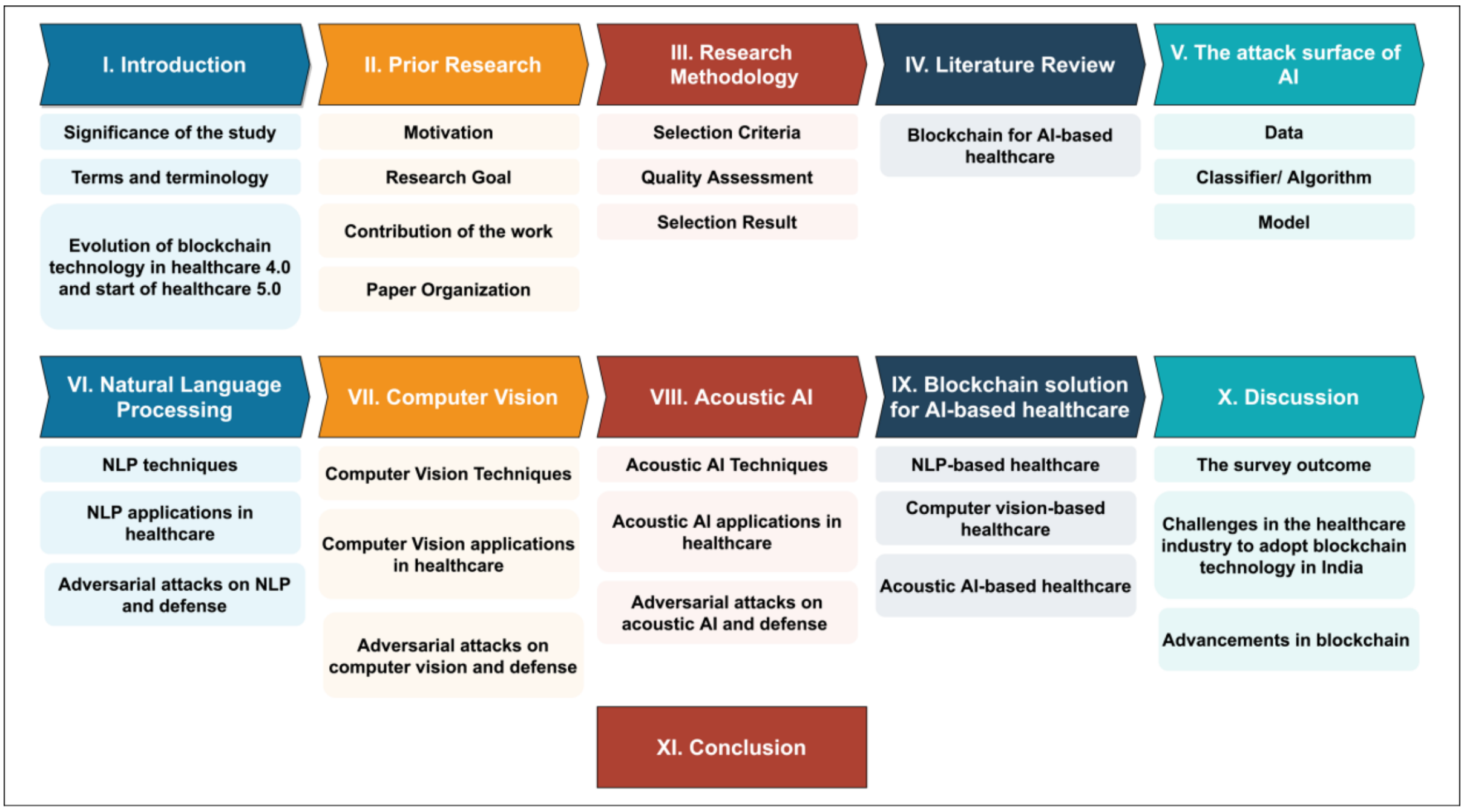

**Figure 5.** Paper Organization.

# III. Research Methodology

The processes or strategies used to locate, select, process, and analyze information on a topic are referred to as research methodology. The authors conducted a literature study using the systematic review approach to answer the research questions targeting the robustness of the healthcare system. This technique is split into the following areas by the authors: selection criteria, quality assessment, and selection result.

## A. Selection Criteria

To find relevant documents, authors have mostly used the Scopus and Web of Science databases. A specific query is created to obtain the documents through multiple database searches. Table 3 depicts the keyword selection process using the PIOC (Population, Intervention, Outcome, Context) method. Moreover, Table 4 represents the search string, i.e., the query used to retrieve the number of documents using the keywords listed in Table 3.

**Table 2.** Keyword Selection.

| Parameter | Meaning | Keywords Used |
|---|---|---|
| **Population** | It is a field of application. | “Healthcare” OR “Medical” |
| **Intervention** | It refers to the software methodology. | “Blockchain” AND (“Artificial Intelligence” OR “Machine Learning” OR “NLP” OR “Computer Vision” OR “Acoustic AI” OR “Federated Learning” OR “Explainable AI”) |
| **Outcome** | It should be related to important aspects, such as increased accuracy, robustness, and trustworthiness. | “Adversarial Attack” OR “Security Issues” or “Privacy Issues” |
| **Context** | It refers to situations in which the solution is carried out. | “Data” OR “Classifier” OR “Algorithm” OR “Model” OR “Text” OR “Image” OR “Video” OR “Sound” OR “Audio” |

**Table 3.** Database Source and Query Executed.

| Database | Search Query | No of Documents |
|---|---|---|
| **Scopus** | (TITLE-ABS KEY (“Healthcare” OR “Medical”) AND TITLE ABS-KEY (“Blockchain”) AND TITLE-ABS-KEY (“Artificial Intelligence” OR “Machine Learning” OR “Federated Learning" OR "Explainable AI" OR "NLP" OR "Computer Vision" OR "Acoustic AI")) | **299** |
| **WoS** | ("Healthcare” OR “Medical") AND ("Blockchain") AND ("Artificial Intelligence” OR “Machine Learning” OR “Federated Learning” OR “Explainable AI” OR “NLP” OR “Computer Vision” OR “Acoustic AI") | **178** |

After conducting a keyword query search in Scopus and WOS, we developed a list of inclusion criteria for selecting research articles and exclusion criteria for rejecting research articles for systematic review. Table 5 shows the inclusion and exclusion criteria that were used for selecting articles for systematic review.

Table 4. Inclusion and Exclusion Criteria.

| Sr.No. | Inclusion Criteria | Exclusion Criteria |
|---|---|---|
| 1 | Published in peer-reviewed journal. | Non-English articles. |
| 2 | Document published after the year 2016. | Book chapters. |
| 3 | Content directly relevant to Blockchain as the solution for AI-based healthcare systems. | Little or no focus on a Blockchain solution for AI-based healthcare systems. |
| 4 | Article answering research questions. | Duplicate articles. |

### B. Quality Assessment

For quality assessment, the authors have considered the following criteria. Only those articles are considered for review which meets these criteria.

- **Application area:** Research emphasizes healthcare applications or medical domain.
- **Objectives:** Articles discussing challenges in AI-based healthcare and mitigating those with Blockchain.
- **Techniques:** Proposed or implemented framework in an article must have blockchain technology integrated with artificial intelligence.
- **Security measures:** Articles must identify features of blockchain and use them to achieve privacy, security, and integrity.

### C. Selection Result

The following Figure 6 represents the selection procedure of relevant articles included for the systematic literature review of “Blockchain for AI-based Healthcare Applications”. At the initial stage, after executing search queries on both Scopus and WoS databases, we received 299 and 178 articles simultaneously. Out of these 477 articles, around 112 articles were found duplicate after screening; hence, authors removed those 112 duplicate articles. In the next step, based on the inclusion criteria that authors have defined, articles were tested for their eligibility. As a result, 342 articles from 365 articles not meeting eligibility criteria were removed from documents collection. Furthermore, out of retained 23 articles, 8 articles never qualified quality assessment that authors have defined to consider the article for Systematic Literature Review. Hence Finally, 15 articles are retained and included in the article collection for conducting a systematic literature review.

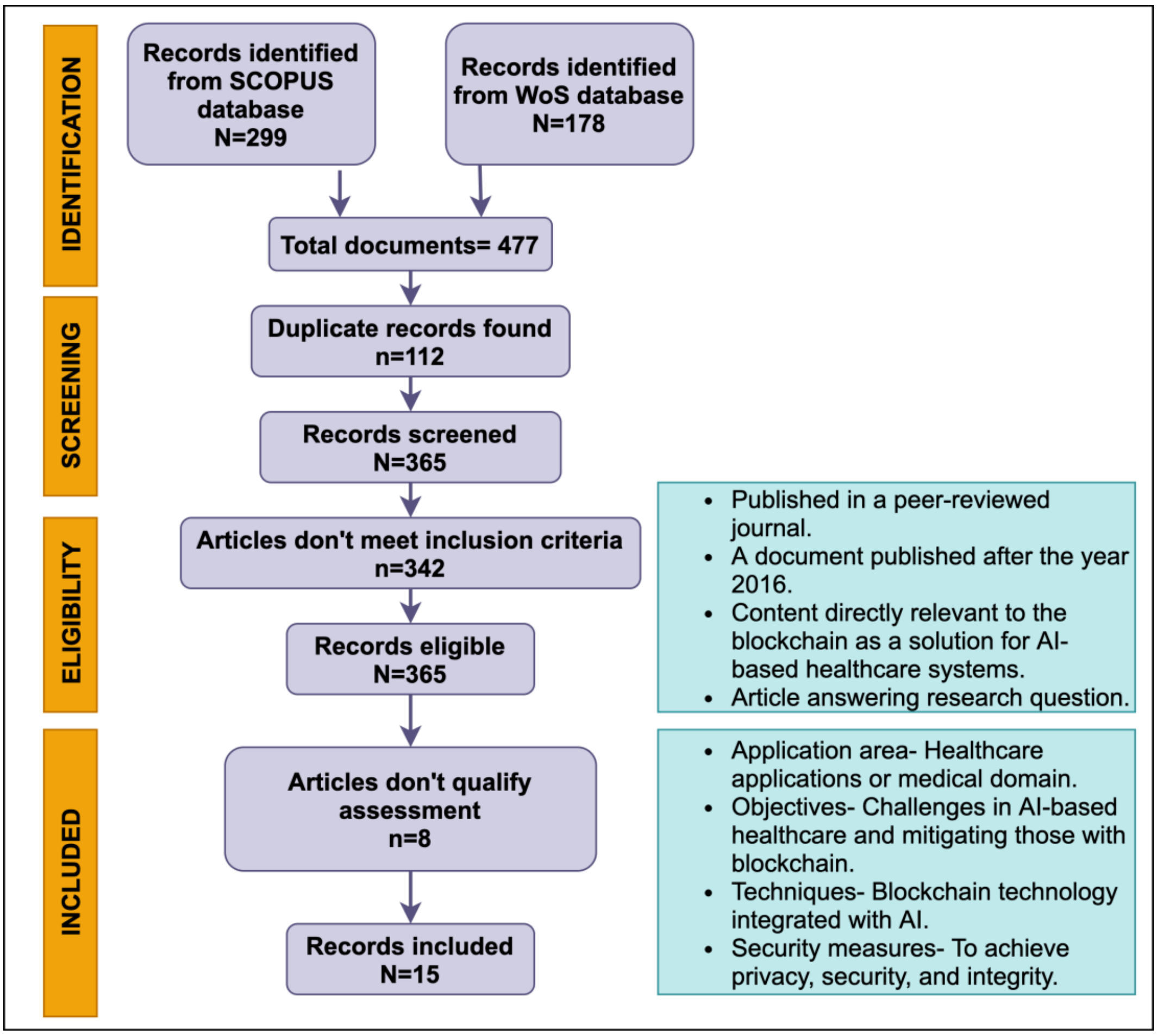

**Figure 6.** Selection Process of Relevant Articles.

# IV. Literature Review

## A. Blockchain for AI-based Healthcare

In analysis of images, deep models have been extensively utilized to tackle several difficulties in medical treatments. When deep learning is trained on a massive quantity of data, it may work more efficiently. Hospitals, diagnostic laboratories, research institutions, and patients may exchange valuable findings and work together to improve the AI model. However, they face challenges in sharing important data with third parties due to privacy and security concerns. Hence, secure data sharing becomes an obstacle in improving the quality of AI-based healthcare systems. Kumar et al. [33] have proposed a strategy that involves sharing local models via the blockchain network, leveraged to collectively develop a global model for improved prediction of lung cancer using CT scan images. As a result, the collectively updated model aids in accurately identifying the patients' disease, resulting in improved therapy. This will avoid actual sharing of data and hence maintains privacy. Through smart contracts, organizations will upload their data over IPFS (Interplanetary File System) and share local gradients. Delegated Proof-of-Stake consensus algorithm is used to train the global model. The trust in data is built over a smart contract, and the hash of the local gradient is maintained in the blockchain. Kim et al. [34] have presented a blockchain algorithm that checks the data that has been refined. The agreement algorithm on the blockchain is HyperPOR. The HyperPOR algorithm functions by confirming the business partner's identity. Then the generation block validates and accomplishes distributed computing, adds sharding technology for protecting PHR(Patient Health Records). Nguyen et al. [35] have proposed an intrusion detection system to protect data transmission in the cyber-physical system of healthcare.

Most of the time, patients have little control over who has access to their medical records and are ignorant of the full worth of the information they possess. Mamoshina et al. [7] have presented AI and Blockchain solutions to speed up biomedical research. Moreover, provide patients with new technologies for controlling and profiting with their personal information and incentives to undertake periodic health checkups. They have proposed Exonum as a permissioned Blockchain framework, wherein patients can sell their health records using tokens. However, this framework does not have control over data once it has been sold to regulators. Jennath et al. [36] proposed a trustworthy Artificial Intelligence model over Blockchain for e-health. An immutable distributed ledger is used to record the provenance of individual permission and the authenticity of data sources to create and train the AI model.

Rahman et al. [37] have used blockchain and off-chain to safeguard from manipulation and illegal access, bringing confidence to the provenance of a dataset and distributed models to protect the privacy and security of the Internet of Health Things (IoHT) data. The insecure central gradient aggregator is replaced with a secure, tamper-proof gradient mining and distributed consensus-based aggregator in the blockchain. The edge training, trust management, and authentication of participating federated nodes, the dissemination of globally or locally trained models, and the identity of edge nodes and their contributed datasets or models are managed by smart contracts. This system provides the complete encryption of both a dataset and a trained model. Puri et al. [38] implemented a decentralized healthcare framework powered by artificial intelligence (AI) that accesses and authenticates Internet of Things (IoT) devices while also instilling confidence and transparency in patient's health records (PHR). The technique is based on AI-enabled smart contracts and the development of a public blockchain network. In addition, the framework detects potentially dangerous IoT nodes in the network. Gupta et al. [39] offer BITS, a unique intelligent TS system based on blockchain. They provide thorough insights into the cloud-based and blockchain-based smart TS frameworks, emphasizing the challenges of security, dependability, confidentiality, and data management. In the case of federated learning, if rogue devices start communicating erroneous local model updates, the global model's accuracy will be skewed. Here blockchain can assure that the local updates in federated learning come from trusted devices. Also, local updates available in blockchain help further verify the accuracy of the learned model. Polap et al. [40] have presented a federated learning methodology that combines decentralized learning with blockchain-based security and a solution for developing intelligent systems with decentralized and locally stored data for the security and privacy of the Internet of Medical Things. This solution can mitigate the model poisoning attacks. Also, [41]stated a way for training a global model cooperatively utilizing blockchain technology and federated learning while maintaining anonymity in detecting covid-19 patients using CT images.

Technological advances, such as distributed learning, provide a road ahead, but they are plagued by a lack of openness, reducing trust in the data utilized for analysis. To solve these challenges, Zerka et al.[42] has projected that Chained Distributed Machine Learning C-DistriM, a novel distributed learning that blends sequential distributed learning with a blockchain-based framework, would be developed in medical imaging. Blockchain is used to record the immutable history of computation and protect from model poisoning. After training, it encrypted the local models and uploaded them on the cloud simultaneously, removing all local copies of the model. Access to the cloud is then restricted through the smart contract for unauthorized users. Kuo et al. [43] have proposed the ExplorerChain framework, which combines two important technologies, online machine learning and a blockchain without a centralized authority, to learn a predictive model across institutions in a distributed architecture, i.e., without the need for patient-level data sharing or a central coordinating node.

Many published deep learning systems lack clarity about model validation and testing outcomes. Blockchain technology might be a viable answer to these challenges, operating as a decentralized, secure, and trustworthy shared ledger for data management and offering tracking and accountability for testing results reporting. Schmetterer et al. [44] deployed a blockchain-based AI platform to establish data transmission, model transfer, and model testing in the real world across three sites in Singapore and China as a proof of concept. Using retinal images taken from distinct multiethnic populations from different countries, researchers set out to create and test deep learning algorithms to identify myopic macular degeneration and extreme myopia. They leveraged a blockchain-enabled AI infrastructure that helps secure, persistent, and verifiable data transmission, model transfer, and transparency in the diagnostic performance of deep learning algorithms. However, it does not maintain the privacy of data.

Khan et al. [45] explored a wireless capsule endoscopy frame-based automated method for detecting stomach infections. A blockchain-based technique is used in a convolutional neural network model to secure the network for precise identification of stomach ailments such as ulcers and bleeding. Each layer comprises an additional block that keeps certain information to resist any tempering and modification attacks. Pilozzi et al. [46] states that artificial intelligence (AI) technologies, particularly natural language processing (NLP), have shown to be an effective tool for classifying the emotion and tonality of texts, like those of social media posts. These approaches could be used to investigate the public perception of Alzheimer's disease. Incorporating secure and decentralized data transfer and storage methods like blockchain will give patients greater control over their data. It will help relieve most of the insecurities of unknowingly disclosing personal information to an entity that may discriminate against the patient.

The following Figure 7 gives an overview of existing work done in blockchain for AI-based healthcare. It represents the type of blockchain used for the different modalities of data. There are three types of blockchain i.e. Public, Private, and Consortium Blockchain. Ethereum is mostly preferred for public blockchain and hyperledger fabric, hyperledger sawtooth for private blockchain. We have focused on text, image, and audio modality of data in healthcare. It has been identified that no specific work done in blockchain for acoustic AI.

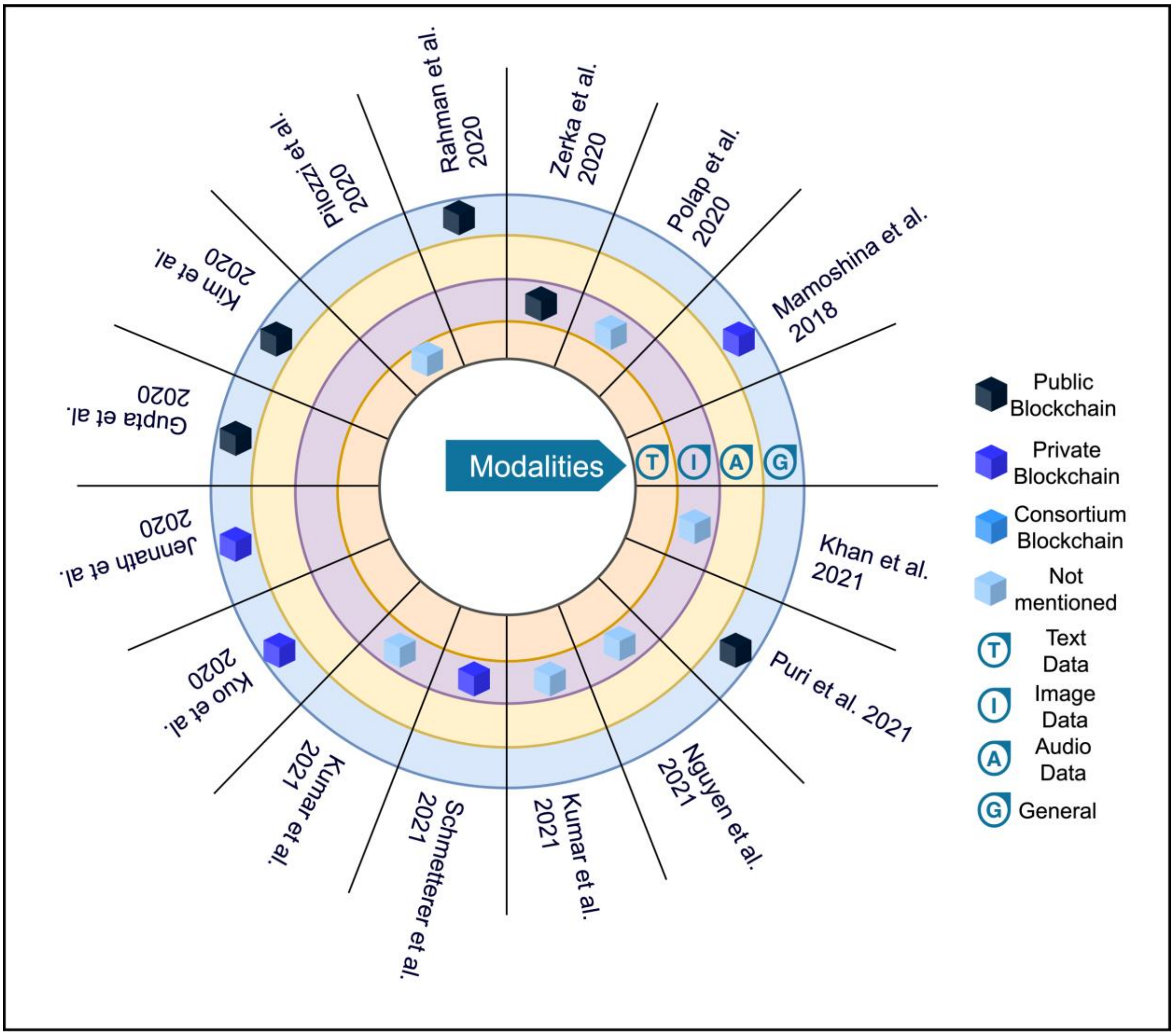

**Figure 7.** Blockchain for AI Healthcare: Types of Blockchain, Data

The following Table 6 provides the overview of existing work done in blockchain for AI healthcare systems considering their goal, implementation and performance evaluation, limitations, and future scope.

**Table 5.** Existing Literature in Blockchain for AI Healthcare.

| Sr. No. | Ref. | Year | Goal | Implementation | A novel blockchain framework | Smart contract | Consensus algorithm | Type of blockchain | Performance evaluation | Limitations | Future scope |
|---|---|---|---|---|---|---|---|---|---|---|---|
| 1 | [7] | 2018 | To propose a blockchain-based secure and distributed personal data marketplace for deep learning technologies in healthcare. | × | ✓ | ✓ | ✓ | ✓ | × | The proposed solution does not protect data leakage once it has been sold. | Blockchain featured AI data marketplace can create advanced research and treatment capabilities in healthcare. |
| 2 | [40] | 2020 | Security and privacy of the Internet of Medical Things with blockchain. | In progress | × | × | × | × | × | No proper architecture is suggested for blockchain in federated learning. | Experimentation can be expanded for different types of data and classifiers. |
| 3 | [42] | 2020 | To promote transparency and confidence in multicentric research and eventually to speed AI adoption. | ✓ | ✓ | ✓ | ✓ | ✓ | ✓ | Not mentioned | To explore how C-DistriM behaves when fraudulent collaborators are deliberately introduced to the network. |
| 4 | [37] | 2020 | To protect the privacy and security of Internet of Health Things (IoHT) data. | ✓ | × | ✓ | ✓ | ✓ | ✓ | GPU computation memory may be a problem for professional cloud providers' trusted execution environment enclaves. | To improve enclave computing in the future. |
| 5 | [46] | 2020 | To review technologies like AI and blockchain that can overcome alzheimer's disease stigma | × | × | × | × | × | × | Not mentioned | AI and blockchain can prove invaluable in fighting alzheimer's disease stigma. |
| 6 | [34] | 2020 | To employ artificial intelligence blockchain algorithms to enable secure PHR data verification and accurate medical data verification in medical institutions | ✓ | ✓ | ✓ | ✓ | ✓ | ✓ | Blockchain technology is in its early stages. | The post-commercial aspect should be evaluated to make the system safer and easier for individuals to use. |
| 7 | [39] | 2020 | To resolve security, privacy, and trust issues in telesurgery. | ✓ | ✓ | ✓ | × | ✓ | ✓ | 5G can introduce a delay in communication. | To identify the challenges and issues in the real-life implementation of the proposed framework. |
| 8 | [36] | 2020 | To address the security and privacy of existing e-Health systems. | ✓ | ✓ | ✓ | ✓ | ✓ | ✓ | The system incorporates limited data management capabilities. | Patient mobility and access permission strategies need to be addressed in the future. |
| 9 | [43] | 2020 | To provide an alternative solution based on a decentralized approach for healthcare AI models. | ✓ | ✓ | ✓ | ✓ | ✓ | ✓ | There seem to be many conflicting views and misunderstandings regarding blockchain technology and its possible benefits in healthcare and genomics, as there are many unanswered issues. | To explore the future of blockchain technology in healthcare and genomics applications. |
| 10 | [33] | 2021 | To maintain privacy while allowing an exchange of information to detect lung cancer using CT images. | ✓ | × | ✓ | ✓ | × | ✓ | Not mentioned | Not mentioned |
| 11 | [44] | 2021 | To develop an Algorithm with the management of a variety of datasets from many nations and centers. | ✓ | × | ✓ | ✓ | ✓ | ✓ | Due to the lack of several images with specific lesions, the developed algorithm can only detect the presence or an absence of myopic macular degeneration and not its categories.<br>It does not maintain the privacy of data. | Widespread adoption of blockchain with deep learning might have far-reaching impacts on AI research in medicine. |
| 12 | [41] | 2021 | To address the issue of data sharing among hospitals like privacy and security in case of the training model for detecting covid-19 patients using CT images. | ✓ | × | ✓ | ✓ | × | ✓ | Not mentioned | Not mentioned |
| 13 | [35] | 2021 | To build intrusion detection system using blockchain-enabled data transmission and classification algorithm in healthcare. | ✓ | × | × | × | × | ✓ | Not mentioned | The system performance can be increased with the use of hyper-parameter tuning techniques and a learning rate scheduler. |
| 14 | [38] | 2021 | To resolve single-point failure, security, privacy, and non-transparency issues with the data in remote patient monitoring. | ✓ | ✓ | ✓ | ✓ | ✓ | ✓ | Not mentioned | It is possible to research trustworthy AI using the suggested framework to improve the system's reliability and build more lightweight algorithms to reduce energy and gas costs. |
| 15 | [45] | 2021 | To address the patient information privacy and time-consuming, costly inspection of stomach abnormalities. | ✓ | ✓ | ✓ | × | × | ✓ | Not mentioned | Secure CNN may be made more secure by utilizing various hashing algorithms, and sophisticated integration of layer ledger blocks with convolution neural network. |

## V. The Attack Surface of Artificial Intelligence

Machine learning is a data processing technique that automates the development of analytical models. It is a subfield of artificial intelligence focused on the principle that computers can learn from data, recognize patterns, and make decisions with little to no human involvement. These tasks required obtaining validated data using which classifiers are trained, and after successful training, the model is deployed. It might proceed with retraining and feedback loops for performance improvement. Figure 8 puts the light on different phases involved in the successful deployment of the AI model. It starts with collecting data and prepare it for training by looking for bias or label it. Then, based on the requirement of application classifier is either developed or selected from the existing one. A classifier is trained for the acquired dataset and can be further improved by adjusting parameters. The trained model is deployed at the end, which will proceed with retraining for future enhancement.

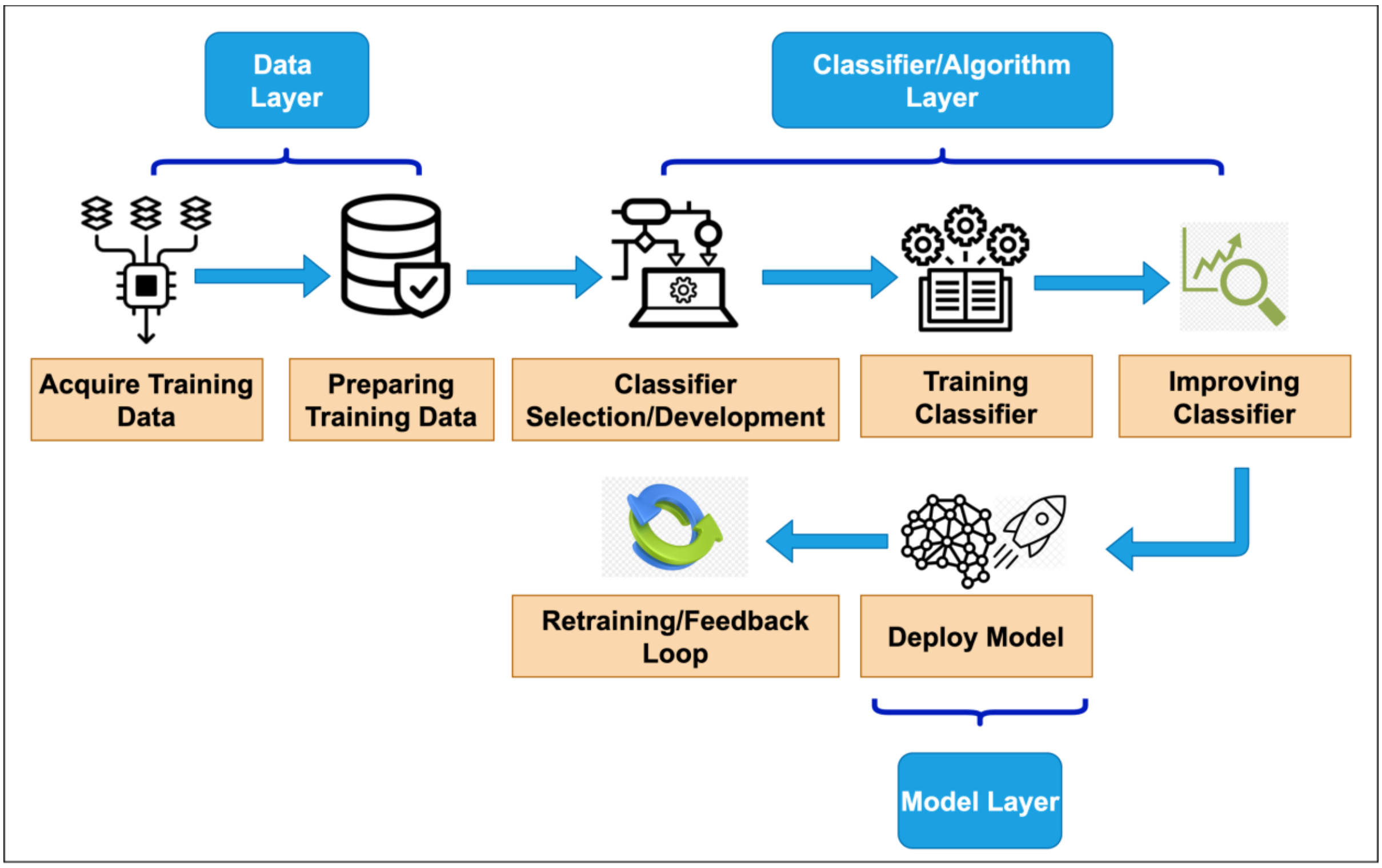

**Figure 8.** Phases Involved in Deploying AI Model.

An attack on a device or an information system is any action to reveal, change, disable, damage, capture, or collect information by exploiting the vulnerabilities available in the system. Maintaining the privacy of sensitive data or process, i.e., confidentiality, making data or process available without tampering, i.e., integrity and allowing data or process to be available to an intended user at any time, i.e., availability is the basic security requirement for any system. Henceforth this CIA triad applies to the phases mentioned above for deploying the ML model securely. In addition, three vitals need to be protected to prevent the malfunctioning of the trained model. Following Figure 9 focuses on the attack surface of artificial intelligence. Data, classifier/ algorithm, and learned model are the targeted areas for imposing attacks in AI-based systems by any adversary.

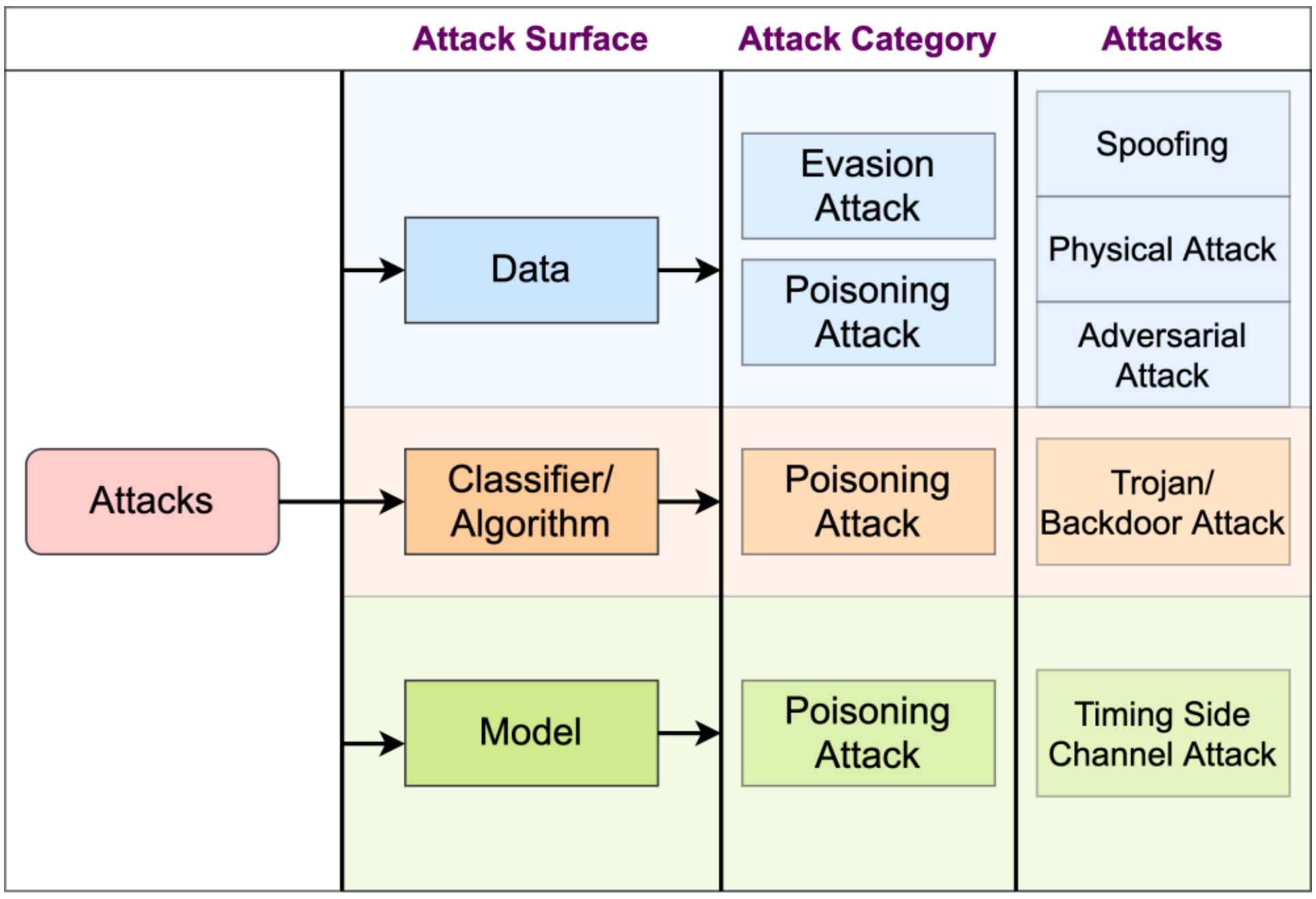

**Figure 9**. Attack Surface of AI.

### A. Data

The most critical component of artificial intelligence is data. Any model cannot be trained without data, and all current technological development will be for naught. There is a big investment of money only to collect as much specific data as possible. Targeting data for the attack has a drastic impact on AI-based systems. Data can be compromised either at the training phase or testing phase by harnessing the extreme sensitivity of AI models to small changes in the input, which exploit abnormal behavior of the model, known as poisoning and evasion attack, respectively. Spoofing [47] can promote these attacks. It is a cyber-attack wherein an adversary attempts to mislead other computer networks by impersonating a genuine entity using a computer, device, or network. Malicious opponents frequently lack access to the training phase of the model. They create adversarial input to deceive a classifier or elude detection from a neural network during the testing phase. These can be either physical or digital kind of attacks. We are focusing on digital kinds of attacks for this study. A digital approach imparts tiny perturbation in the input directly. Here the attacker can exploit the targeted system without being detected by the intrusion detection system. Evasion attacks can also lead to concept drift [48]. Potential attackers can also gain access to the dataset used for training and contaminate the dataset with adversarial samples, known as poisoning attacks. Adversarial attacks in AI-based healthcare have the potential to harm human beings, as discussed in further sections.

### B. Classifier/ Algorithm

#### 1. Trojan/Backdoor Attack

A trojan attack compromises the genuine model by inserting a backdoor into the neural network, activated with a specific pattern in the testing sample. A trojan attack targets the origin of vulnerabilities in a neural network like data, algorithm, and program or hardware. It will modify the network with a compromised dataset [49]. At the time of the training phase, neural trojans are injected into the network [50]. Trojan attacks are different that of adversarial attacks, although both take place at the training phase only. In the case of an adversarial attack, it does not force the neural network to modify itself; rather, it only affects the result. However, in a trojan attack, poisoned input samples make the network modify itself to work accurately for benign input samples, and it will malfunction only when triggered by a trojan. It is difficult for a user to identify the trojan attack [51]. A Stealthy Poisoning Attack (SPA) is depending on a Generalized Adversary Network (GAN), which can lead to a trojan attack [52]. Badnet [53] is also an example of a neural trojan attack. The following Figure 10 depicts the scenario of a trojan attack.

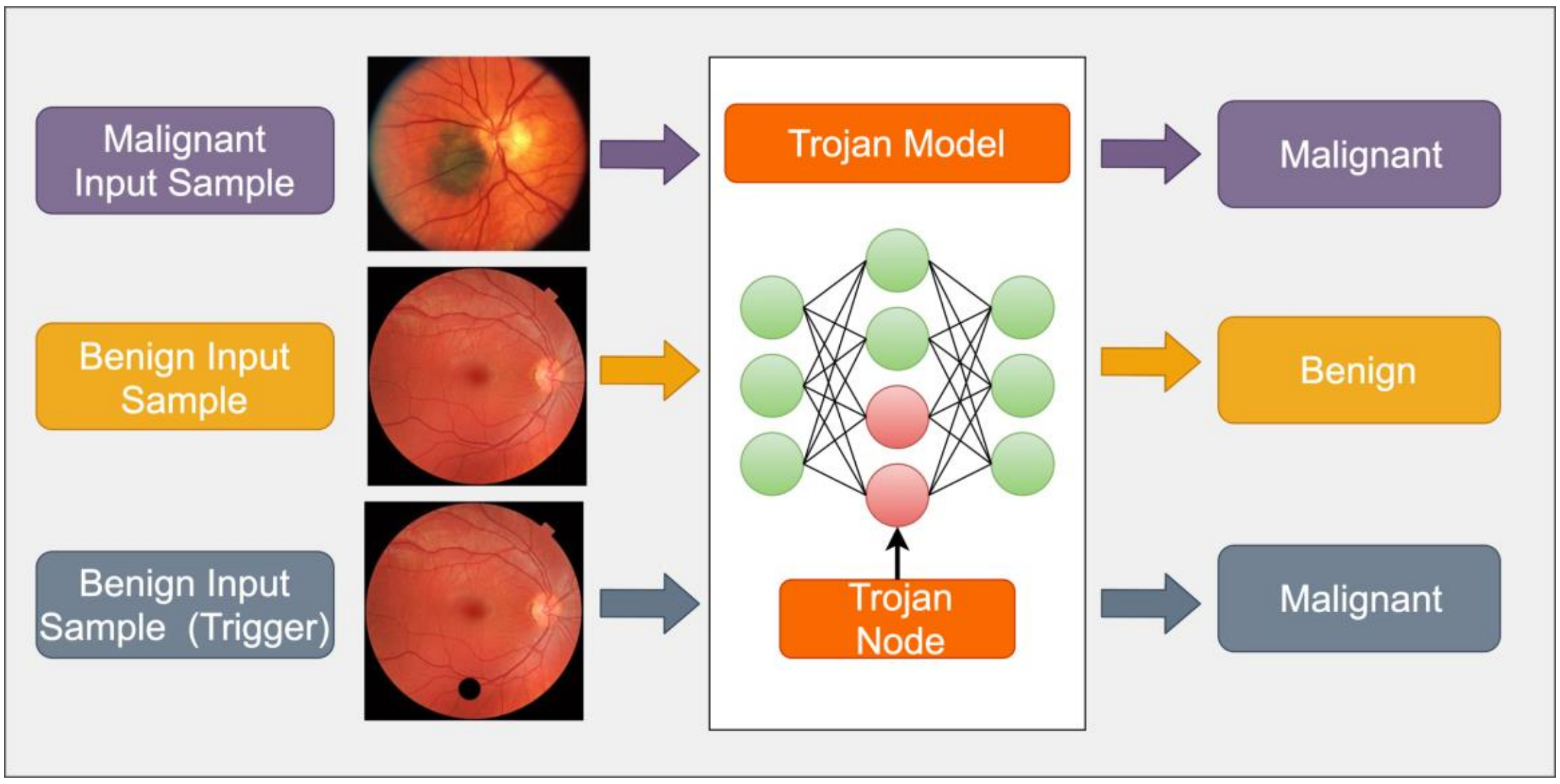

**Figure 10.** Trojan Attack.

### C. Model

#### 1. Timing Side-Channel Attack

The characteristics of neural networks for having different times for execution based on the depth of the network makes it vulnerable to attacks like timing side channel. By observing the time required for generating output by the model, an adversary can conclude with the number of layers, i.e., the depth of the neural network. This adversary uses a regressor trained using different execution times along with the respective number of layers in the network. This information is then utilized to mimic substitute models with similar functionalities to the original network [54]. With the help of memory and timing side-channel attacks, information of the CNN model can be leaked using reverse engineering of structure and weights. The memory access patterns explode the vital features of a neural network like the total number of layers, size of the layer, and their dependencies [55].

## VI. Natural Language Processing (NLP)

### A. NLP Techniques

#### 1. Named Entity Recognition

The most fundamental method in NLP is retrieving entities from text. It emphasizes the key topics and connections in the text. Named Entity Recognition (NER) extracts entities from text such as persons, places, organizations, dates, etc. Grammar rules and supervised models are commonly used.

#### 2. Sentiment Analysis

Sentiment analysis is the most used approach in NLP. Sentiment analysis is particularly effective when individuals express their thoughts and feedback, such as customer surveys, reviews, and social media comments. The most basic result of sentiment analysis is a three-point scale: positive/negative/neutral. The result can be a numerical score grouped into as many categories as the user desires in more sophisticated instances.

#### 3. Text Summarization

This NLP method aids in the summarization of lengthy passages of text. Text summarization is commonly utilized in situations such as news articles and research papers. Extraction and abstraction are two major techniques for text summarization. Extraction methods generate a summary by extracting pieces of writing. Abstraction approaches provide a summary by producing new text that expresses the essence of the original material.

#### 4. Aspect Mining

Aspect mining identifies the various perspectives in a text. It retrieves entire information from the text that can be used in combination with sentiment analysis. Part-of-speech tagging is one of the simplest ways of aspect mining. When aspect mining and sentiment analysis are applied to a text, the outcome reveals the whole intention of the text.

#### 5. Topic Modeling

One of the more difficult approaches for identifying natural concepts in the text is topic modeling. The fact that topic modeling is an unsupervised approach is a significant benefit. There is no need for model training or a labeled training dataset.

### B. NLP Applications in Healthcare

There are many inputs to Clinical Decision Support Systems (CDDS) like semi-structured data, such as XML documents or two-column laboratory results, structured data such as Electronic Health Records (EHR) and narrative text, patients' clinical findings, radiology records, and operative notes are examples of unstructured data. Several solutions have been proposed that use NLP techniques with unstructured data as input to support clinical decisions, especially to compute and automate diagnoses or treatments. NLP provides adequate mechanisms for automated extraction of important facts from free text, which CDSS uses to generate the results and recommendations provided to healthcare professionals to assist them in making the best decisions [56]–[58]. The clinical notes, including the patients' health history, are crucial assets for resolving sensitive clinical problems difficult to access from other EHR components, like lab results. NLP makes it easy to retrieve meaningful features from clinical notes that can be used to build machine learning models. Unified Medical Language System (UMLS) resources and clinical notes become effective and valuable tools with NLP for predicting mortality in diabetic patients in the critical care environment. However, we will need more research databases and patient cohorts to test the model [59]. To determine sentiment in a repository of narrative hospital admission and discharge statements, a sentiment scoring algorithm, also known as opinion mining, has been used [60]. Radiology reports are also in unstructured form. NLP makes it possible to identify and extract important features from those reports and convert them into manageable computer formats [61]. Based solely on the content of clinical notes, the supervised machine learning-based NLP approach to identify the medical subdomain of a clinical note can assist clinicians in directing patients' unresolved problems to appropriate health professionals and experts promptly [62]. NLP can be used to investigate viral transmission, ocular manifestation, and course of treatment for ophthalmology patient care by analyzing COVID-19 ophthalmology-related articles [63]. NLP has the potential to revolutionize procedure-heavy sub-disciplines like gastroenterology by analyzing a tremendous amount of free-text narrative medical reports [64]. To identify pediatric emergency department patients with a high suspicion of Kawasaki isease (KD), KD-NLP made significant improvements to clinician manual chart analysis [65]. Adverse drug effects (ADE) and medication-related knowledge can be extracted using a well-designed hybrid NLP framework, used in real-world applications to facilitate ADE-related scientific and drug decisions [66]. The chatbot system will be fed with information about different diseases, and it will be able to understand the user question and respond appropriately using NLP [67]. There is evidence for wide acceptance of deep learning for clinical NLP due to the increasing volume of medical data [68]. NLP has been used in recent research to classify diseases and disorders that are hard to diagnose using only clinical gestalt. Information Retrieval (IR) takes less time and effort when NLP-based solutions are used, which ultimately foster the treatment [69], [70].

### C. Adversarial Attacks on NLP and Defense

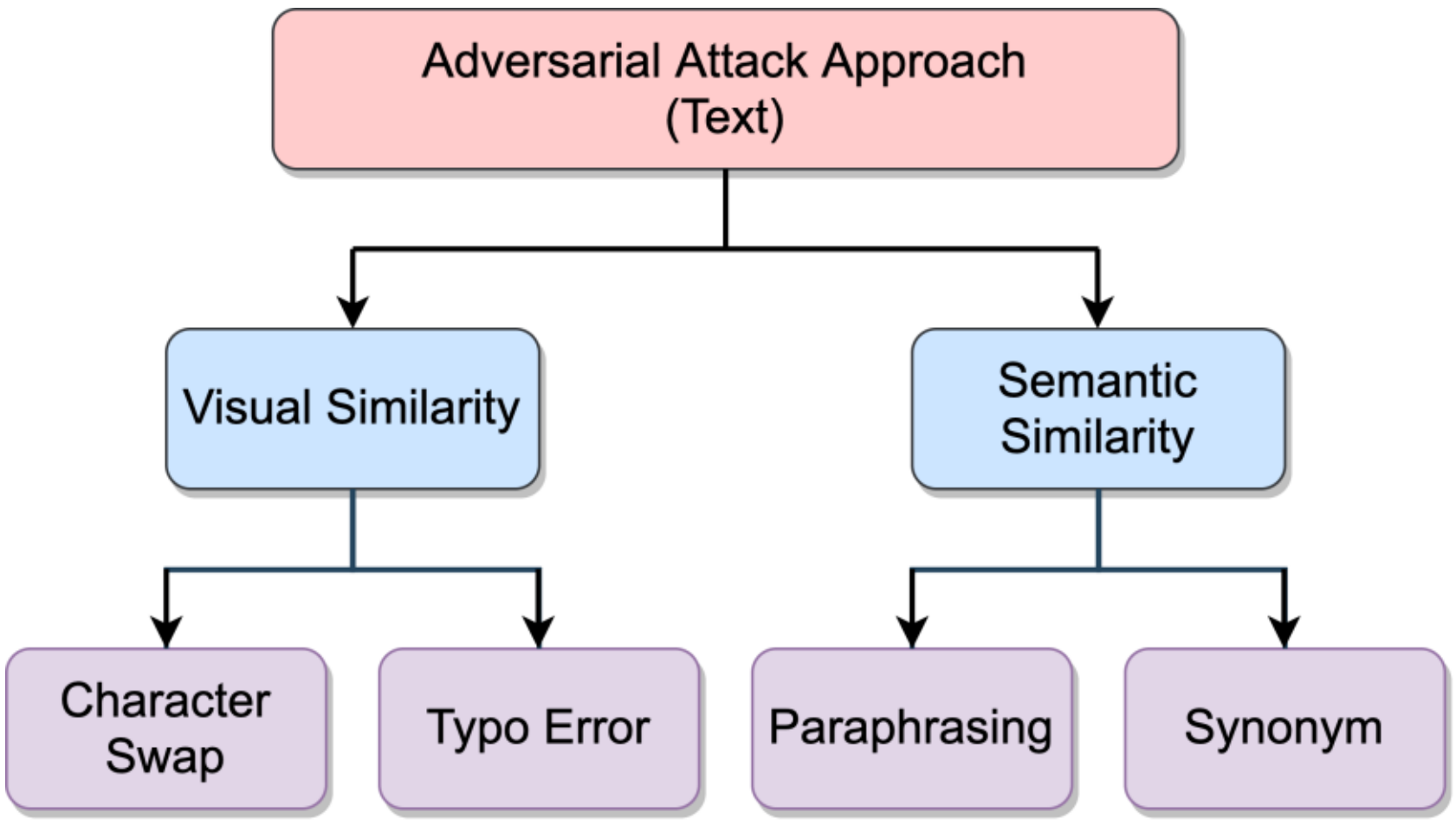

**Figure 11**. Taxonomy of Adversarial Attack on Text.

Adversarial text is a perturbation in the text regarding semantics, syntax, and visual similarity, which will mislead NLP. The Figure 11 depicts the methods for generating adversarial text. Adversarial examples in text can be a tiny attempt to modify minimum characters and create genuine typing errors comparable to those made by humans to affect the model's prediction. This attack keeps the sequence of appearance of text very close to the original one. The attacks like DeepWordBug [71], hotflip [72], and textbugger [73] come under this approach of attack. Another approach to generating adversarial text depends on paraphrasing the original text. This attack will generate the semantical equivalence of the original text, but the model output will be different for the original and paraphrased text. Alzantot [74], bae [75], Bert-attack [76], IGA [77], PWWS [78], textbugger [73], textfooler [79] are some attacks which generates adversarial text under mentioned approach. Textbugger is the combination of both approaches for targeting adversarial attacks on NLP.

The Figure 12 illustrates the toxic effect of an adversarial attack on NLP-based Healthcare applications. Just replacing the words with their synonym while maintaining the semantic of the text can fool the NLP. The wrong prediction of the severity of the disease ultimately leads to wrong treatment and puts lives at risk. Table 7 gives overview of adversarial attacks on text.

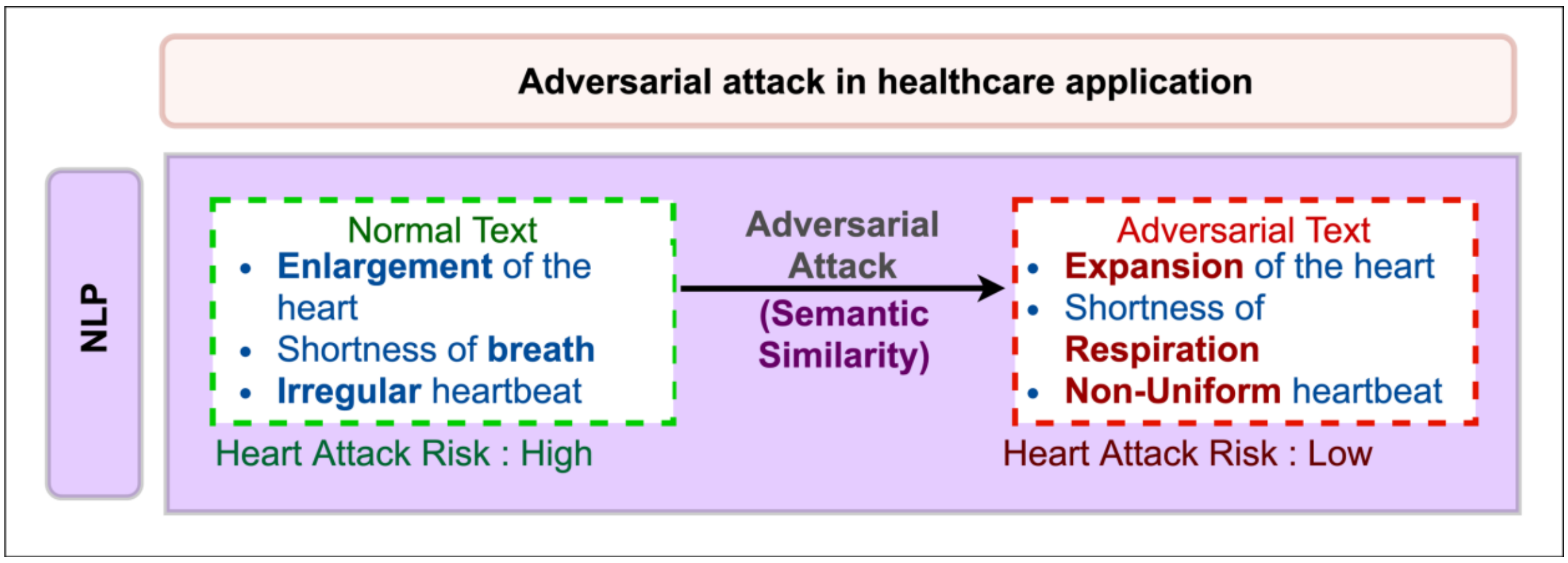

**Figure 12.** Adversarial Attack on NLP-based Healthcare Application.

**Table 6.** Adversarial Attacks on NLP.

| Approach | Attack | Threat Model | Perturbation | Target Model |
|---|---|---|---|---|
| Visual Similarity | DeepWordBug | Black Box | Character Level | Word-LSTM, Char-CNN model |
| | Hotflip | White Box | Character Level, Word Level | CNN |
| | Textbugger | White Box/ Black Box | Character/ Word Level | LR, CNN, LSTM |
| Semantic Similarity | Alzantot | Black Box | Word Level | LSTM |
| | Bae | Black Box | Sentence Level | Word-LSTM, Word-CNN |
| | Bert-attack | Black Box | Word Level | CNN |
| | IGA | Black Box | Word Level | Word-CNN, LSTN, Bi-LSTM |
| | PWWS | Black Box | Word Level | Word-CNN, Char-CNN, LSTN, Bi-LSTM |
| | Textfooler | Black Box | Word Level | Word-CNN, Char-CNN, LSTN |

The following Table 8 depicts the defense techniques for adversarial attacks on the NLP model, along with their limitations. It has been observed that adversarial training is the most robust defense technique for a particular adversarial attack on the NLP model.

**Table 7.** Defense Techniques for Adversarial Attacks on NLP.

| Adversarial Attack | Defense Technique | Accuracy | Limitation |
|---|---|---|---|
| DeepWordBug | Adversarial Training | 62% | A new class of attacks makes adversarial training-based defense vulnerable. |
| HotFlip | Adversarial Training | 69.32% | The discrete property of texts makes it difficult to broaden the categories of attacks covered during training. |
| TextBugger | Adversarial Training | >75% | A new class of attacks makes adversarial training-based defense vulnerable. |
| IGA | Synonym Encoding Method (SEM) | >50% & <87% depending on dataset and network model | Works better for only Greedy Search Algorithm, PWWS, Genetic Algorithm, and IGA attacks |
| PWWS | Adversarial Training | 80% | Defense is limited to PWWS attack |

# VII. Computer Vision

## A. Computer Vision Techniques

### 1. Image Classification

A collection of images labeled with a specific category is available, and then to forecast these categories for a new set of test images and assess the accuracy of those predictions. This job has several difficulties, including perspective variation, size variation, intra-class variance, image distortion, image occlusion, lighting conditions, and backdrop clutter.

### 2. Object Detection

Defining objects inside pictures often entails producing bounding boxes and labels for each item. This procedure varies from the classification/localization job in that classification and localization are applied to many items rather than simply a single dominating object. There are only two types of object classification: object bounding boxes and non-object bounding boxes.

### 3. Object Tracking

The practice of following a particular object of interest, or several objects in each scene, is referred to as object tracking. Historically, it has been used in video and real-world interfaces where inspections are performed after an initial object identification.

### 4. Semantic Segmentation

The technique of segmentation, which splits whole pictures into pixel groupings that may subsequently be labeled and categorized, is essential to computer vision. Semantic segmentation attempts to grasp the semantic role of each pixel in a picture. We must also define the bounds of each item. As a result, unlike categorization, we require detailed pixel-by-pixel predictions from our models.

### 5. Instance Segmentation

Instance segmentation goes beyond semantic segmentation by segmenting various instances of classes. There is usually a picture with a single item in the center in classification, and the objective is to identify that image. However, to segment instances, we must do considerably more complicated activities. We observe complex scenes with many overlapping items and diverse backgrounds, and it not just classify these things but recognizes their borders, distinctions, and relationships to each other.

## B. Computer Vision Applications in Healthcare

Intelligent intervention using a brain-like structure aids in the understanding and analysis of different forms of dynamic data by using cutting-edge technologies such as deep learning and computer vision [80]. Computer vision is a scientific form of deep learning, where it detects objects from captured series of videos and images. Deep learning algorithms known as convolutional neural networks (CNNs) are designed to process image data and assign importance to various aspects to distinguish one image. CNNs have a structural architecture like the connectivity pattern of neurons in the brain. In object-classification activities, state-of-the-art computer vision accuracy outperformed as compared to human accuracy. The National Cancer Institute's National Lung Screening Trial (NLST) found that lung cancer screening with low dose computed tomography (CT) has reduced mortality by 20% [81]. The ImageNet Large-Scale Visual Recognition Challenge (ILSVRC) brought together a large group of deep learning researchers who competed and collaborated to develop strategies for a variety of computer vision tasks [82]. Computer vision has a longstanding experience of allowing computers to interpret visual imagery meaningfully. Identifying the type of an object in an image, the position of present objects, and both type and location simultaneously are referred to as object classification, localization, and detection, respectively [83].

Advances in computer vision and deep learning have improved the effectiveness of smart monitoring. An innovative computer vision-aided deep learning derived posture monitoring system is introduced for predicting Generalized Anxiety Disorder (GAD) oriented physical abnormalities in a person from their workplace [84]. At the same time, patient mobilization occurs early and often reduces the risk of post-intensive care syndrome and long-term cognitive dysfunction. In an adult ICU, computer vision algorithms identify patient mobilization activities like moving the patient into and out of bed and into and out of a chair [85]. Deep convolutional neural networks provide a significant opportunity as a diagnostic tool for otologic prognosis based on otoscopic ear images [86]. Researchers combined computer vision approaches with deep learning neural network techniques to create a comprehensive image processing model to predict human embryo viability [87]. Automated herbal medications based on tongue images are possible with deep learning to investigate the relevance of tongue with herbal medication [88]. Using a deep learning based automated platform to test Diabetic Retinopathy (DR) from color fundus images may offer an alternative approach for reducing medication errors. A DL-based automated tool has significant advantages in cutting down screening rates, increasing healthcare coverage, and ensuring early treatment [89], [90]. The confidence-aware anomaly detection (CAAD) model, considered a feature extractor, an anomaly detection module, and a confidence predictor module for detecting viral pneumonia from chest X-rays, may be of considerable use for large-scale screening and infection control [91]. To detect hip fractures from pelvic x-ray is also possible with a computer vision algorithm [92]. The revolutionary CNN method for detecting esophageal cancer, including squamous cell carcinoma and adenocarcinoma, will analyze recorded endoscopic images quickly and accurately [93]. The automatic identification of Intracranial Hemorrhage (ICH) and its forms from non-contrast head CT scan scans is possible by deep learning algorithms [94]. A diagnosis from chest CT images is feasible, resulting in a deep learning algorithm in a fast and automated diagnostic method [95]. A novel visual SLAM algorithm can also track and locate robots in dynamic situations to reduce the chance of doctor-patient cross-infection and prevents the growth of the COVID-19 [96].

### C. Adversarial Attacks on Computer Vision and Defense

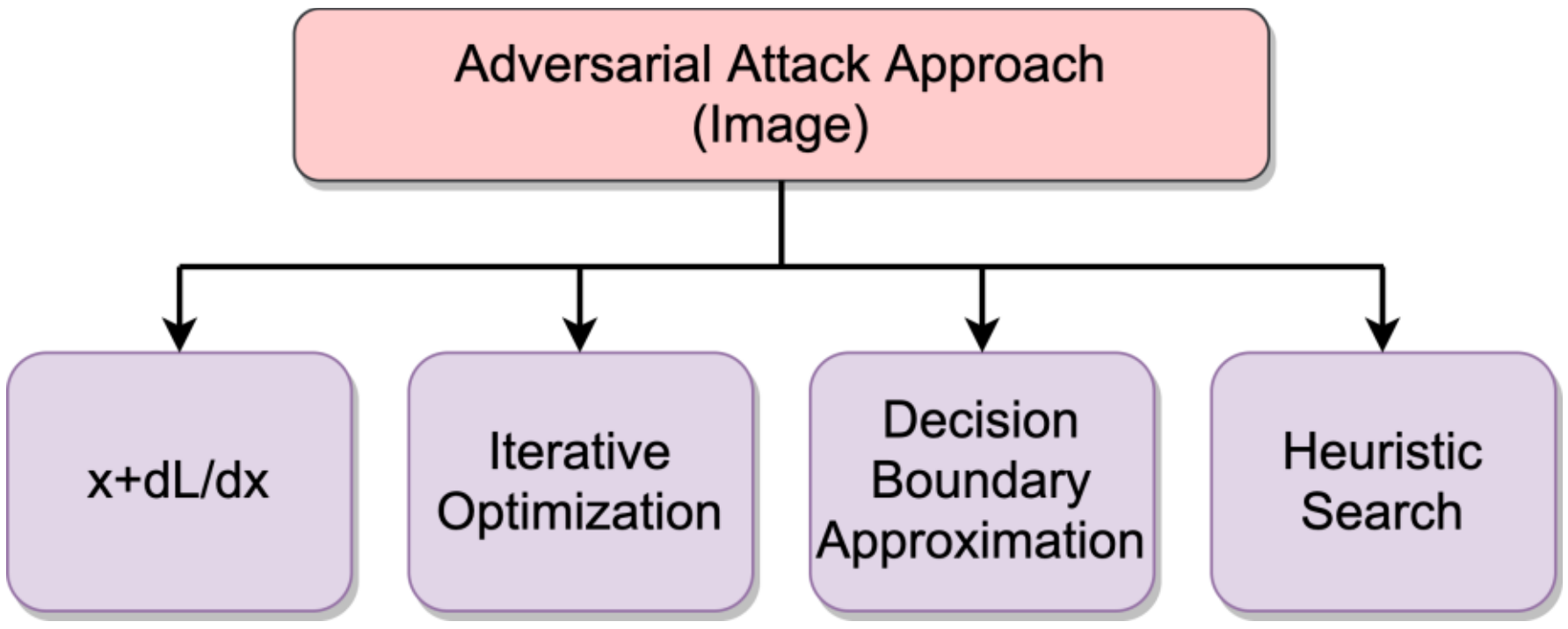

**Figure 13.** Taxonomy for Adversarial Attack on Image.

Adversarial images are those wherein pixels are intentionally perturbed to confound and fool models; simultaneously, they seem innocuous and benign to human eyes. Adversarial images mislead DNN as a minute perturbation in the input makes DNN vulnerable. The Figure 13 represents the approaches for an adversarial attack on an image. Some approaches for attack depend on first order derivative by computing the derivative of the loss concerning input for targeting an increase in the loss, e.g., FGSM [97], BIM [98], R+FGSM [99]. Other attacks depend on an incremental optimization procedure on multiple goal objective functions, allowing an attacker to include additional hostile criteria into the targeted function, e.g., L-BFGS [100], C&W [101], ATN [102], stAdv [103], Deepfool, Universal Adversarial Perturbations (UAP). An adversary may benefit from the transferability property of adversarial samples to attack a blackbox model by targeting a substitute model trained using a labeled dataset of the black-box model, e.g., substitute black box attack [104]. Another black-box approach is based on crawling closer to the decision boundary within adversarial input and non-adversarial input, e.g., boundary attack [105].

Following Figure 14 reflects the adversarial effect of adding some perturbation in the original hand x-ray image. Perturb image is identical to the original image from human perception, but the AI model interprets the wrong result after processing. For example, the adversarial attack changes the prediction from a hairline fracture to normal. Table 9 gives details of adversarial attacks on image.

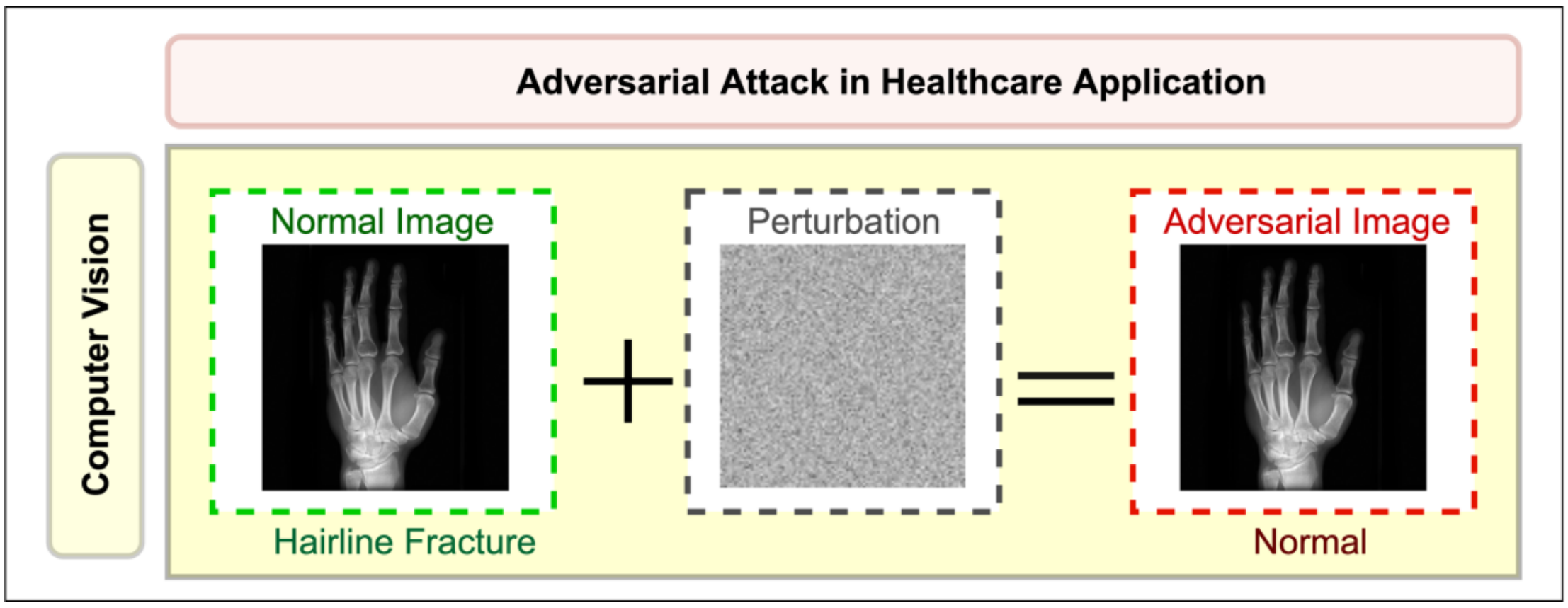

**Figure 14.** Adversarial Attack on Computer Vision-based Healthcare Application.

Table 8. Adversarial Attacks in Computer Vision.

| Approach | Attack | Distance Metric | Threat Model | Objective |
|---|---|---|---|---|
| Derivative Function | FGSM | $L_\infty$ | White Box | Targeted Attack |
| | BIM | $L_\infty$ | White Box | Non-Targeted Attack |
| | R+FGSM | $L_\infty$ | White Box | Targeted Attack |
| Iterative Optimization | L-BFGS | $L_\infty$ | White Box | Targeted Attack |
| | C&W | $L_0$, $L_2$, and $L_\infty$ | White Box | Targeted Attack |
| | ATN | $L_\infty$ | White Box | Targeted Attack |
| | stAdv | NA (Spatial Location-Based) | White Box | Targeted Attack |
| | UAP | $L_1$, $L_\infty$ | White Box | Targeted Attack |
| | Deepfool | $L_2$ | White Box | Non-Targeted Attack |
| Decision Boundary Approximation | Boundary Attack | NA (Decision Based) | Black Box | Targeted/ Non-Targeted Attack |
| Heuristic Search | Substitute Model | NA | Black Box | Targeted Attack |

The following Table 10 highlights the advantages and disadvantages of defense techniques designed to mitigate a different adversarial attack on computer vision models. It has been identified that the defense technique can handle only a particular kind of attack with adversarial training and other techniques also have certain limitations. Hence to frame a global solution to mitigate all kinds of adversarial attacks is a challenging task.

**Table 9.** Defense Techniques for Adversarial Attack on Computer Vision.

| Defense Technique | Advantages | Disadvantages |
|---|---|---|
| Adversarial training [106]<br>1. FGSM adversarial training<br>2. Adversarial logit pairing<br>3. PGD adversarial training<br>4. Ensemble adversarial training | It achieves state-of-the-art accuracy on several benchmarks. | It will fail for the data outside the training set/new attack. |
| Randomization [106]<br>1. Random input transformation<br>2. Random noising<br>3. Random feature pruning | Best results for black-box and grey-box settings. | Not effective in white-box setting. |
| Denoising [107] | It can mitigate the C&W attack. | EAD and CW2 can bypass the input squeezing system with increasing |
| 1. Conventional input rectification<br>2. GAN-based input cleansing<br>3. Auto encoder-based input denoising (MagNet)<br>4. Feature denoising | | adversary strength. Adaptive white-box CW2 attack can easily defeat APE-GAN. MagNet is vulnerable to the transferable adversarial samples generated by the CW2 attack. HGD is compromised by a PGD adversary under a white box setting. |
| Provable defenses [107]<br>1. Semidefinite programming-based certificated defense<br>2. Dual approach-based provable defense<br>3. Distributional robustness certification | It maintains a certain accuracy under a well-defined class of attacks. | Scalability has been a common problem shared by most certificated defenses. |
| QR Code [108] | Alteration in the dataset is easily identified. | The storage requirement increases linearly with the number of images in the dataset. |

# VIII. Acoustic AI

## A. Acoustic AI Techniques

### 1. Selective Noise Cancelling

Artificial intelligence noise canceling technology improves headset microphones by removing background noise, resulting in crystal-clear online audio conversation. A processor with AI-enhanced profiles eliminates more than 50 million different forms of background noise while maintaining speech harmonics.

### 2. HI-Fi Audio Reconstruction

Recently, there has been a surge of interest in utilizing deep neural networks to perform this up sampling on raw audio waveforms. To train neural networks to infer additional time-domain samples from an audio signal, equivalent to the picture super-resolution issue in which individual audio samples are analogous to pixels. It is possible to train a network to generate realistic audio by teaching how a normal recording sound.

### 3. Analog Audio Emulation

It involves calculating the complicated interactions of non-linear analog audio components. It investigates how a deep neural network (DNN) can learn the lengthy temporal dependencies that define these effect units with the possibility of matching nonlinearities within the audio effects using convolutional, recurrent, and fully connected layers. It investigates linear and nonlinear time-varying emulation as a content-based transformation without explicitly finding the time-variant solution.

### 4. Speech Processing

The computer receives input data of sound vibrations in speech recognition. This is accomplished by employing an analog to digital converter, which transforms the sound waves into a digital format that the computer understands. Advanced speech recognition in AI also includes AI voice recognition, in which the computer can recognize the voice of a specific speaker.

### 5. Improved Spatial Simulation

Improved spatial simulation is used for binaural processing and reverb. Binaural processing in hearing equipment typically employs device communication to recognize and improve the loudest speech signal in the environment. Reverb is caused by sound reflecting and echoing off walls, floors, ceilings, or other surfaces.

## B. Acoustic AI Applications in Healthcare

The "Firefly" app is an application framework and underlying software development kit (SDK), which uses advanced digital signal processing (DSP) technology and artificial intelligence (AI) algorithms to detect thorough sleep cycles, respiration rate, snoring, and obstructive sleep apnea (OSA) patterns, which can be used to measure the precise respiratory rate of a human in bed via smartphones. This application combines active sonar with passive acoustic analysis [109]. The spectral analysis of acoustic signals is used to measure feature vectors and validated a series of machine learning methods to provide the most efficient identification of cardiac valve defects based on heart sounds. The convolutional neural network has shown to be an effective method for increasing efficiency [110]. A new method of automated sound processing based on neural networks (NNs) has been introduced in a framework that captures respiratory sounds using an electronic stethoscope. It can recognize four types of auscultatory sounds: wheezes, rhonchi, fine and coarse crackles which lead to a reduction in human errors, while auscultation sound perception [111]. Using a deep CNN-RNN model that classifies respiratory sounds based on mel-spectrograms, classification models and techniques were developed to classify breathing sound anomalies (wheeze, crackle) for automatic diagnosis of respiratory and pulmonary diseases proved to be an efficient methodology[112], [113]. Also, researchers have used the audio characteristics of coughs to build classifiers that can differentiate various respiratory disorders in adults.

Furthermore, they have used recent developments in generative adversarial networks to balance and expand a dataset by adding brilliantly constructed synthetic cough samples for each type of serious respiratory disease. CoughGAN creates simulated coughs that reflect significant pulmonary symptoms to aid physicians and predictive algorithms in detecting the early stages of lung disease. Thus, clinicians will establish the right preventative treatment plans and reduce morbidity by detecting advanced respiratory conditions like chronic obstructive pulmonary disease early and accurately [114]. Most of the research has been conducted in AI-based classification of respiratory sound with the use of Noise Masking Recurrent Neural Network (NMRNN) [115], Support Vector Machine (SVM) for pediatric breath sound classification [116], back-propagation neural network model[117], convolutional recurrent neural network (CRNN) with Reinforcement Learning (RL) agent [118] The use of a novel CNN architecture (N-CNN) in conjunction with other CNN architectures (VGG16 and ResNet50) to measure pain in newborns from crying sounds. The findings of the experiments show that using innovative N-CNN for measuring pain in neonates has much therapeutic significance and is a good alternative to the conventional evaluation method [119].

### C. Adversarial Attacks on Acoustic AI and Defense

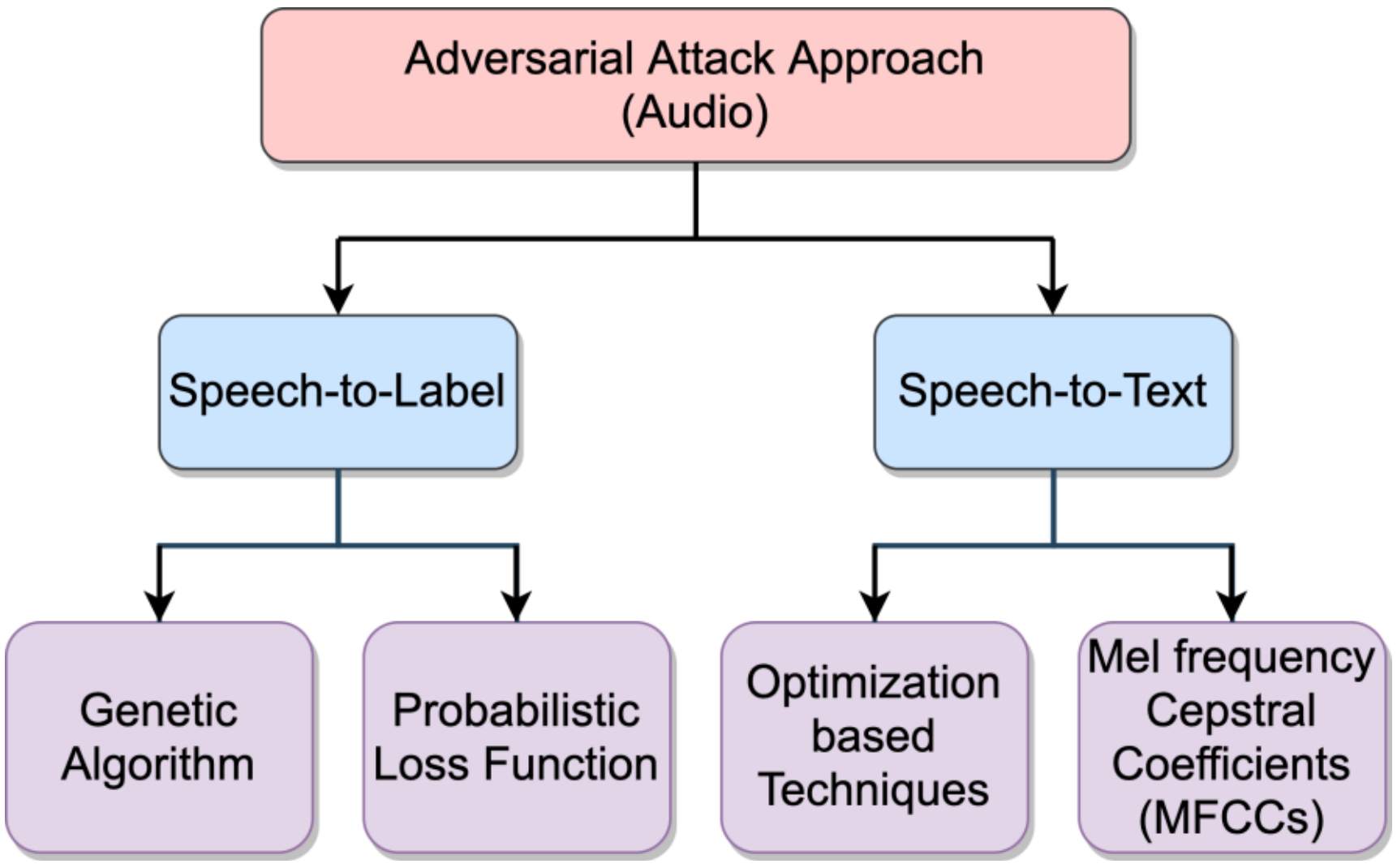

**Figure 15.** Taxonomy for Adversarial Attack on Audio.

Adversarial audio is the sound with perceptible noise, i.e., adversarial perturbations, which can deceive a range of sound classification systems. The Figure 15 represents the taxonomy for an adversarial attack on audio signals. Some attacks target generating an adversarial audio sample that seems very similar to the original one, but the learned model would result in the wrong classification. These attacks come under speech-to-label kind of attacks. The attacker can demonstrate genetic algorithms free from gradients to generate adversarial audio samples just by accepting original audio and desired output label. It adds random noise but keeps it away from human perception [120]. Deceiving the cutting-edge learning systems with undetectable perturbation in audio is made possible by Houdini using a probabilistic loss function [121]. An adversary can also try to force desired output in acoustic processing while converting speech into text. These attacks are known as speech-to-text attacks. Mel Frequency Cepstral Coefficients (MFCC) parameters of input audio signal can be modified, and then it will be reconstructed with reverse MFCC applied to the genuine extracted features [122]. Changing the audio spectrum into the desired transcript can be changed by merely adding a small distortion using optimization-based attacks [123].

The following Figure 16 shows that after processing the audio signal of cough sound, which is perturbed by adding frequency, it has changed the prediction from normal cough to bronchitis. These adversarial attacks spoil the integrity of AI models.

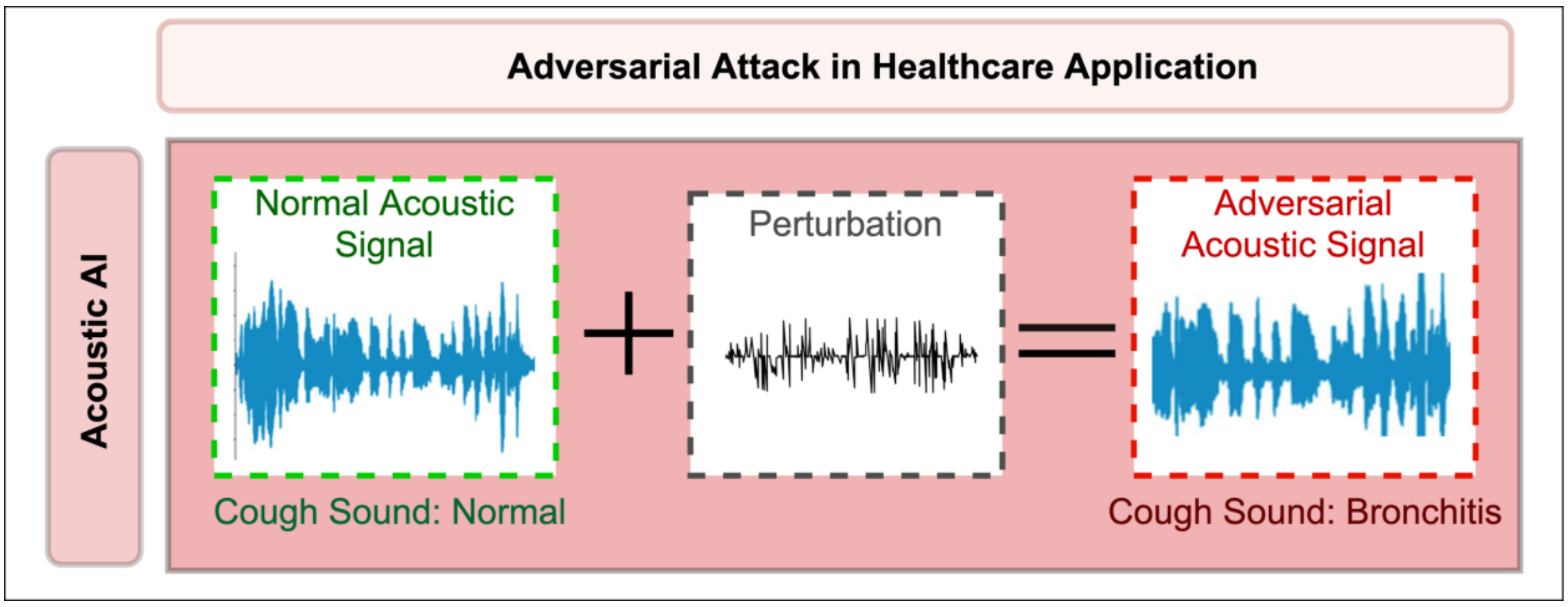

**Figure 16.** Adversarial Attack on Acoustic-based Healthcare Application.

Unfortunately, there are very few studies in the literature on protecting against adversarial audio attacks. Constructing a general and strong defensive mechanism to address this issue is still a work in progress. [123] has developed a proactive defensive system to restore the original input's structure and eliminate the perturbations. General protection against voice recognition systems, which includes code modulation and audio compression. Code modulation combines the G.729 narrow-band vocoder-based audio data compression method with PCM to modify and convert the audio waveform. By utilizing MP3 compression, audio compression eliminates unwanted audio information that surpasses human hearing. [125] has discovered that frame offsets with a blank clip inserted at the beginning of audio can deteriorate adversarial perturbations to normal noise by examining the features of the automatic speech recognition system. Even though this technique can identify and protect the audio adversarial scenario, the perturbations remain as noise, affecting the system's identification accuracy. [126] has deployed two adversarial training designs to protect against adversarial attacks: vanilla and unique similarity-based adversarial training contribution. Adversarial training intends to train on both actual and fictitious data. However, it needs vast adversarial data to train the detector.

## IX. Blockchain Solution for AI-based Healthcare

AI can enhance care delivery, productivity, and efficiency, allowing healthcare organizations to get more and better treatment to many more patients. In addition, AI can help healthcare professionals better experience by concentrating on acute care and decreasing stress. According to the literature we studied, NLP, Computer Vision, and Acoustic AI are three champions in processing health data that includes text, picture, and audio simultaneously and assisting medical professionals in diagnosing and treating patients. Though there are far more benefits of adopting AI in the healthcare industry, certain threats still slow down its acceptance in critical applications of healthcare. Data, classifier/algorithm, and model are the attack surface of AI. We have studied various adversarial attacks independently on text, image, and audio. We came across certain comparative findings for adversarial attacks on NLP, computer vision, and acoustic AI, as mentioned in Table 11.

**Table 10.** Comparative Analysis of adversarial attacks on Computer Vision, NLP, and Acoustic AI.

| Parameter | NLP | Computer Vision | Acoustic AI |
|---|---|---|---|
| **Nature of data** | Discrete nature | Continuous Pixels | Continuous-Time series |
| **Undergone DNN architectures** | Convolution/ Recurrent | Convolution | Recurrent |
| **Complexity of Attack** | Moderate | Easy | Difficult |
| **Perceivability by Human** | Easy | Difficult | Moderate |

Considering the adverse effect of such attacks on healthcare applications, authors have gone through different defense techniques for adversarial attacks. However, each of the defense techniques has its advantages and disadvantages, as discussed in earlier sections. Furthermore, many of the defense techniques are themselves AI-based and very specific to kind of adversarial attack. So yet, no defense strategy has been capable of managing all sorts of adversarial attacks. Hence, we are proposing a blockchain solution to mitigate those attacks.

### A. NLP-based Healthcare

Natural Language Processing (NLP) is an innovative technique for improving information extraction methods where data is discrete. NLP algorithms can be protected with immutable, tamperproof, and trusted blockchain technology by carrying critical data extracted after NLP algorithms, fulfilling the requirement of data provenance and 100% uptime for the system with its distributed nature. As discussed in the above section, NLP suffers from various kinds of adversarial attacks. However, the complexity of those attacks is moderate, and in the case of textual data, adversarial attacks are easily perceivable by human eyes. Hence, we can conclude that the probability of such attacks might be moderate. We have framed blockchain solutions for addressing attack surfaces like data, classifier, and model in NLP.

**Synthesized Framework:**

1. Dataset Building:

In NLP, data can reside on local machines of data owners like doctors, hospitals, laboratories, etc. They can build a peer-to-peer network of blockchain within distributed owners to resolve the need for sufficiently large data for training AI models and at the same time maintain the privacy of data by allowing data owners to not share data directly to the third party. This framework supports off-chain storage of data. This peer-to-peer network offers a direct exchange of services through a proper authentication mechanism. Thousands of machines can be connected without any centralized server. The node in the P2P blockchain network can act as either requester or service provider. Through smart contracts, access control rules can be implied to have authorized and trusted sharing of data. A hash value is generated at each data station and maintained in blockchain to check distributed data integrity. A hash will be recalculated and validated through blockchain at the time of data utilization for training purposes. The following Figure 17 illustrates the dataset building with blockchain for NLP.

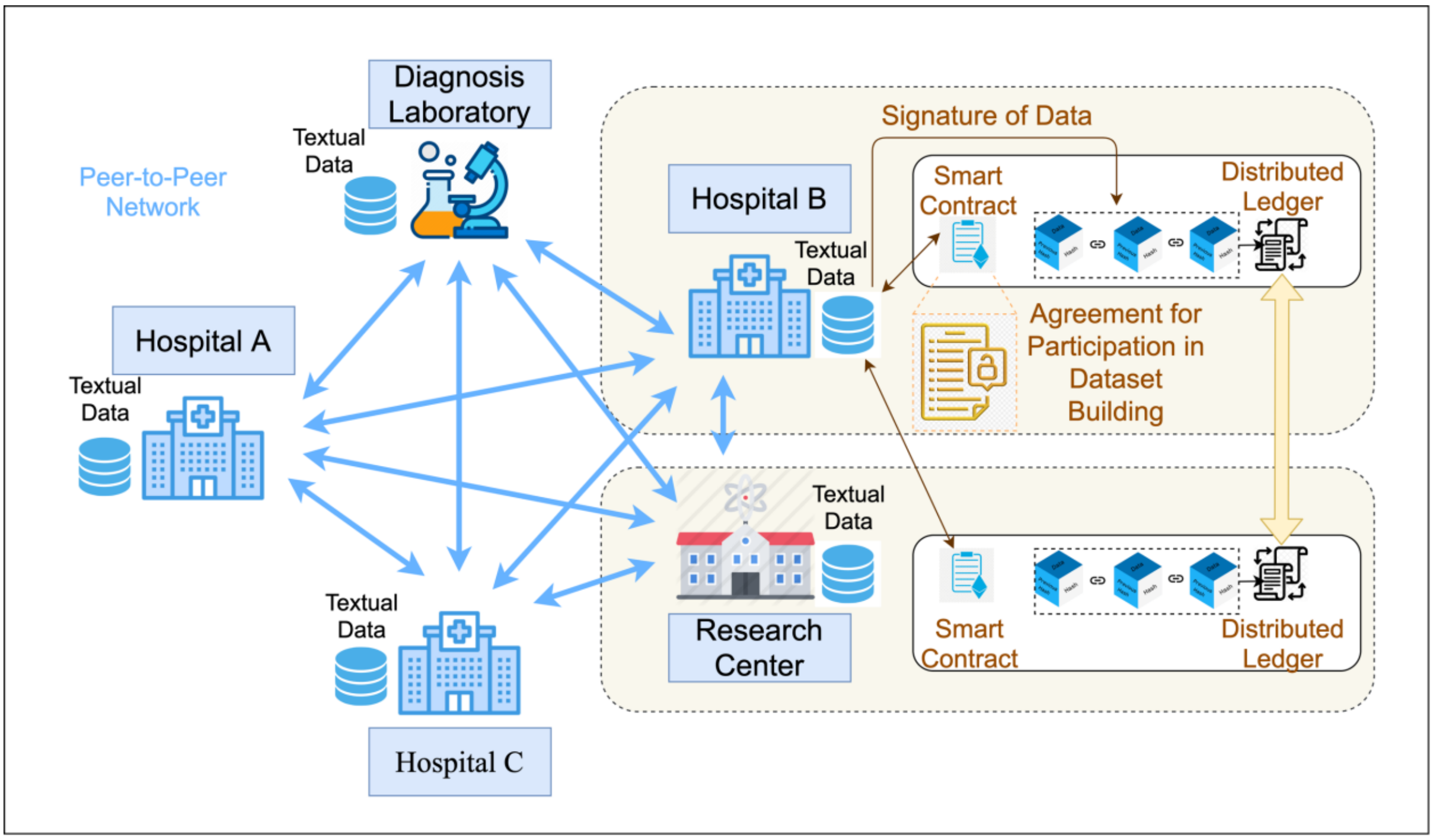

**Figure 17.** Dataset Building with Blockchain in NLP-based Healthcare Applications.

Participants in the blockchain network can be various hospitals, diagnosis laboratories, research centers, etc., with a database of textual data like clinical reports and notes. There will be peer-to-peer connections within each participant, and blockchain will govern their communication. For example, suppose the research center needs access to the dataset from the hospital, then it will initiate a request for data access through a smart contract. Hospitals can respond with an agreement for participation in dataset building, mentioning rules or restrictions for data sharing and usage for training AI-model. At each participant, copy of the distributed ledger is available. The hash value can be generated and stored in the blockchain as its signature for future integrity checks of data for data at each station. In this way, many data stations can contribute to trusted dataset building with blockchain.

2. Training Phase:

Federated learning comes into the picture to train a learning algorithm using the distributed dataset. Federated learning faces challenges like a malfunctioned node, trust in local gradients, and global gradient aggregation. Leveraging blockchain in federated learning helps solve those challenges and protect the model from poisoning attacks. Training can be initiated through a smart contract so that it can check on the participant's legitimacy.

Local gradients at federated nodes will then be communicated through block. It will lock local gradients in the blockchain to avoid tampering and be used in the future for verification. Miners in the blockchain network will validate and generate global gradients using a consensus algorithm. This is how blockchain can bring authenticity to the federated network. Each node embeds extracted features in vector space, and those will be saved in distributed ledger for further use. Word embeddings are a kind of word representation that meaningfully connects a human's grasp of knowledge to a machine's understanding. For example, a set of real numbers can be used as a depiction (a vector). Word embeddings are a dispersed representation of a text in an n-dimensional space that attempts to capture the meanings of the words. The following Figure 18 illustrates the blockchain solution for protecting classifier/ algorithm in NLP.

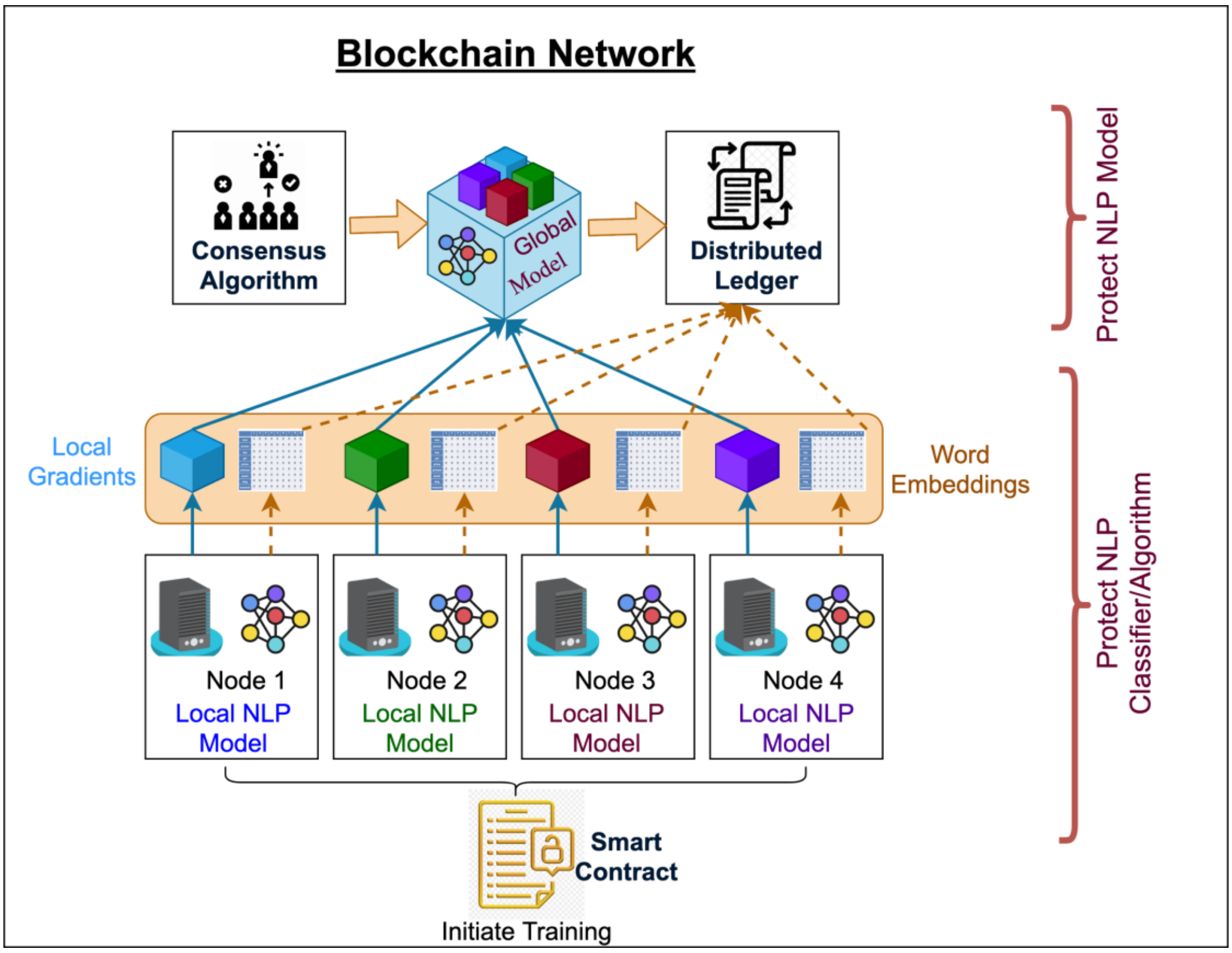

**Figure 18.** Blockchain Solution to Protect Training Phase of NLP.

3. Post Training Phase:

The output of the trained model depends on how genuine the input is in the post-training phase. For adversarial text input, we might expect malfunctioning of the NLP model. By leveraging the blockchain in NLP-based Healthcare, we can try to mitigate certain adversarial attacks. The blockchain version of extracted features from a dataset used for model training plays an important role in discriminating adversarial text. For the given input, word embeddings will be generated through a smart contract. It will look for a similar corpus of word embeddings in the distributed ledger based on synonyms. The resultant extracts are further distributed among miners in the blockchain network. Instead of proof of work, miners will compute the result for assigned features using a trained model. Then results will be shared among miners, and based on a majority of votes, the result will be recorded in the blockchain consensually. This framework will defend the model from the synonym-based adversarial attack on text as described in Figure 19.

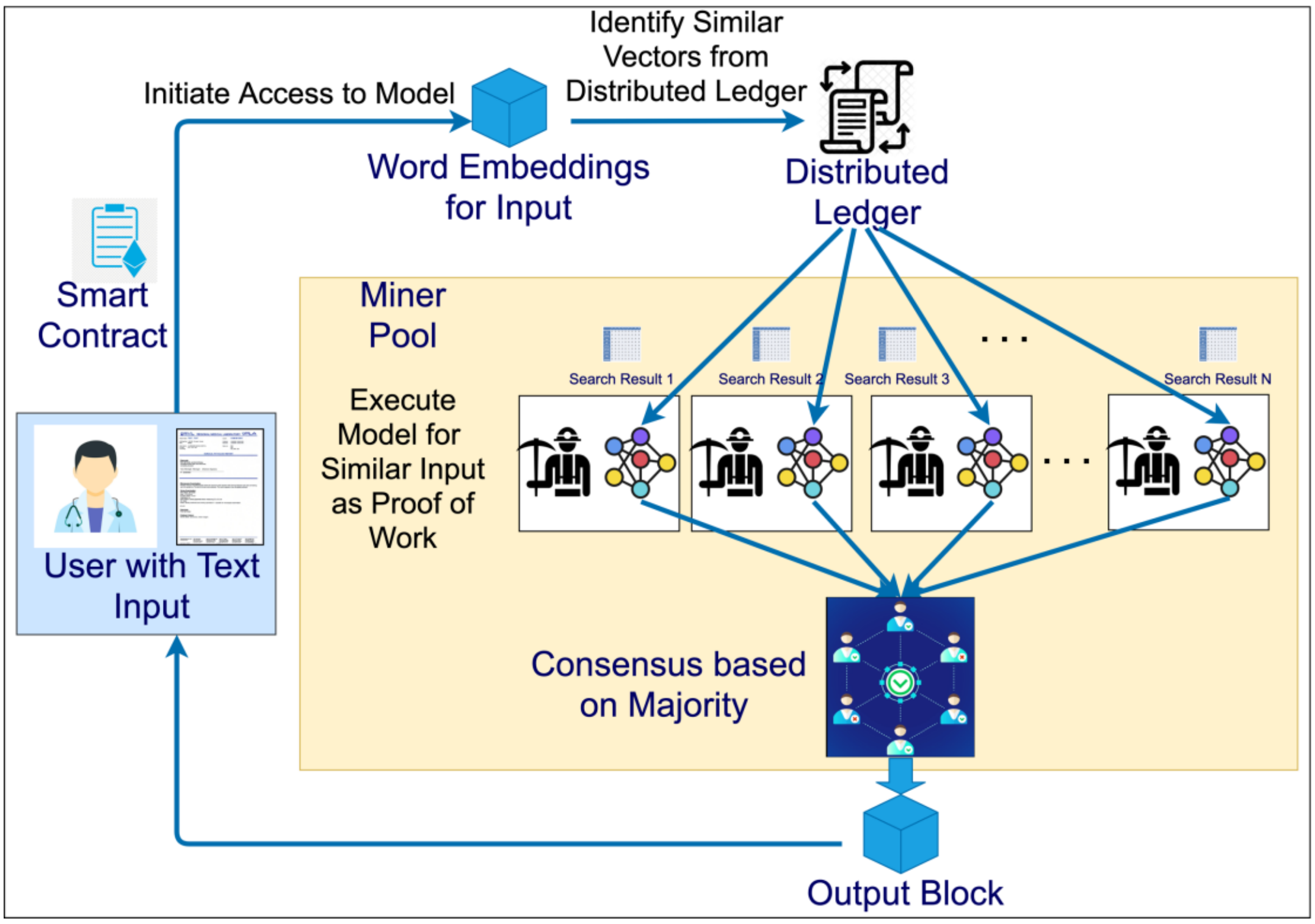

**Figure 19.** Post-training Blockchain Solution for NLP-based Healthcare.

### B. Computer Vision-based Healthcare

Computers must "understand" an image and comprehend its features to take full advantage of images. Computer vision enriches computers with the capabilities to unlock the image data. Computer visions play a vital role in healthcare, as discussed before but at the same time it also prone to malfunction due to potential adversarial attacks. AI-based healthcare applications cannot survive in such an adversarial environment. After studying the adversarial attacks on computer vision, we found that the complexity of adversarial attacks on images is less and also those are non-perceivable. Hence more focus should be on designing preventive measures. Here we are proposing a systematic strategy to handle the adversarial environment in computer vision with blockchain.

**Synthesized Framework:**

1. Dataset Building:

Images are more prone to adversarial attacks, so we frame solutions focusing on preventive measures, as shown in Figure 20. Images will be uploaded to IPFS with the help of blockchain technology. IPFS is a file-sharing technology that may be used to store and distribute big files more effectively. It is based on cryptographic hashes, which can be effortlessly preserved on a blockchain. Cryptographic hashes are used to validate the integrity of images. First, data owners like different hospitals, diagnostic centers request for uploading medical images on IPFS. Smart contracts validate the data owner and upload data on IPFS. For an uploaded set of images, a cryptographic hash is generated at IPFS, which is further stored in blockchain. The research centers which need medical image datasets to train an AI model can request blockchain to access image datasets. Smart contract authenticates the user and provides hash value stored in blockchain for data available in IPFS. Users will approach IPFS with hash value and get access to the image dataset. It will be easy to detect adversarial images with hash values as small changes in images will drastically change in hash generated. This is how the dataset will be secure at IPFS with blockchain.

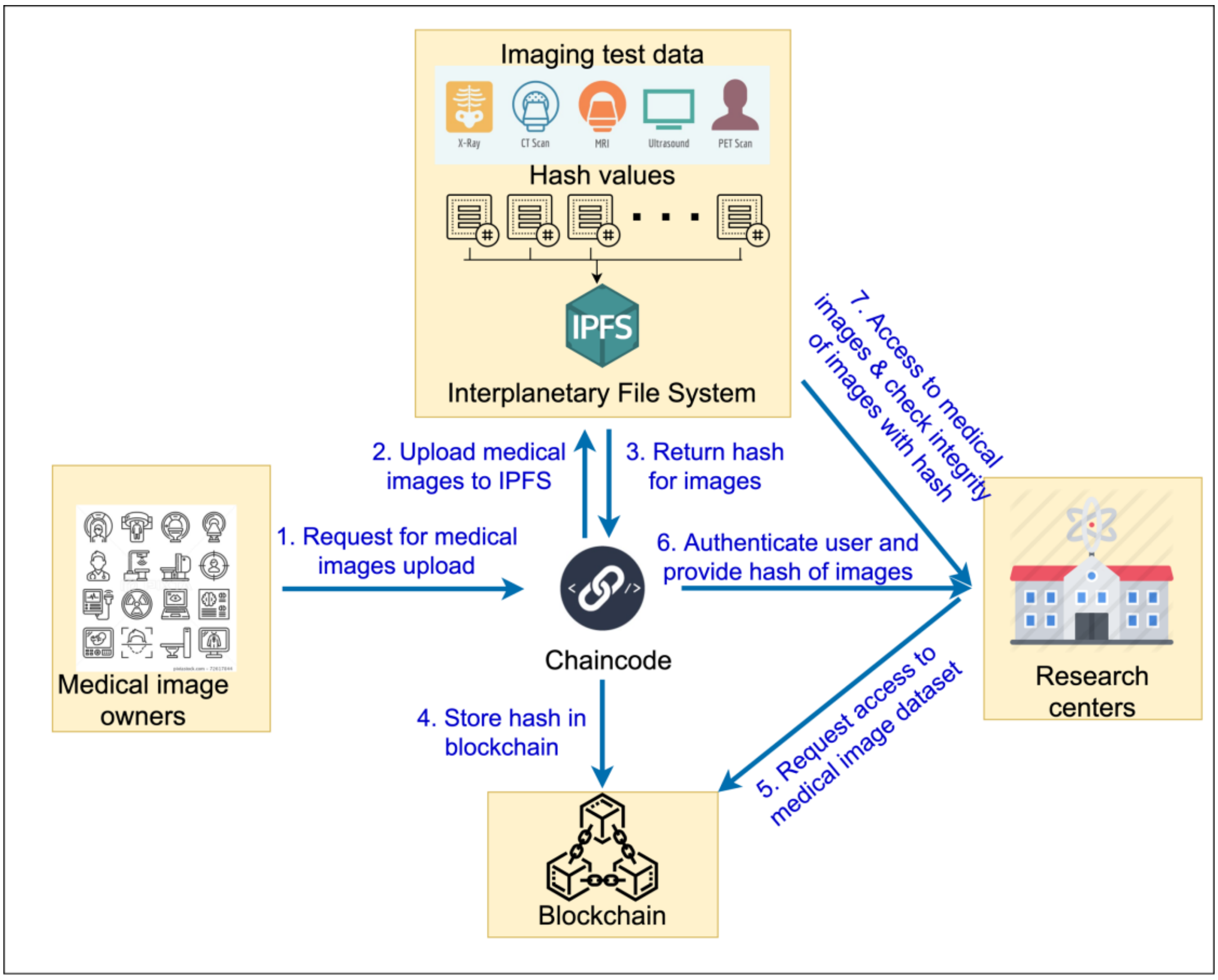

**Figure 20.** Dataset Building in Computer Vision-based Healthcare Applications.

2. Training Phase:

For visual inputs, computer vision provides real-time data. Therefore, we must train our model with image datasets, informing the model of what each input shows. This training phase can be protected with blockchain, as shown in Figure 21. Research centers must initiate the learning process through smart contracts only after proper authentication. Once the model is trained, the extracted features will be stored in blockchain for further reference in feature vector $x = (x_1, x_2, x_3, \ldots, x_n)^T$, where n gives the number of features extracted and T represents transpose operation along with metadata. This framework will protect the training entire training environment for computer vision. Thus, we will have a tamper-proof record of features extracted by the learning process.

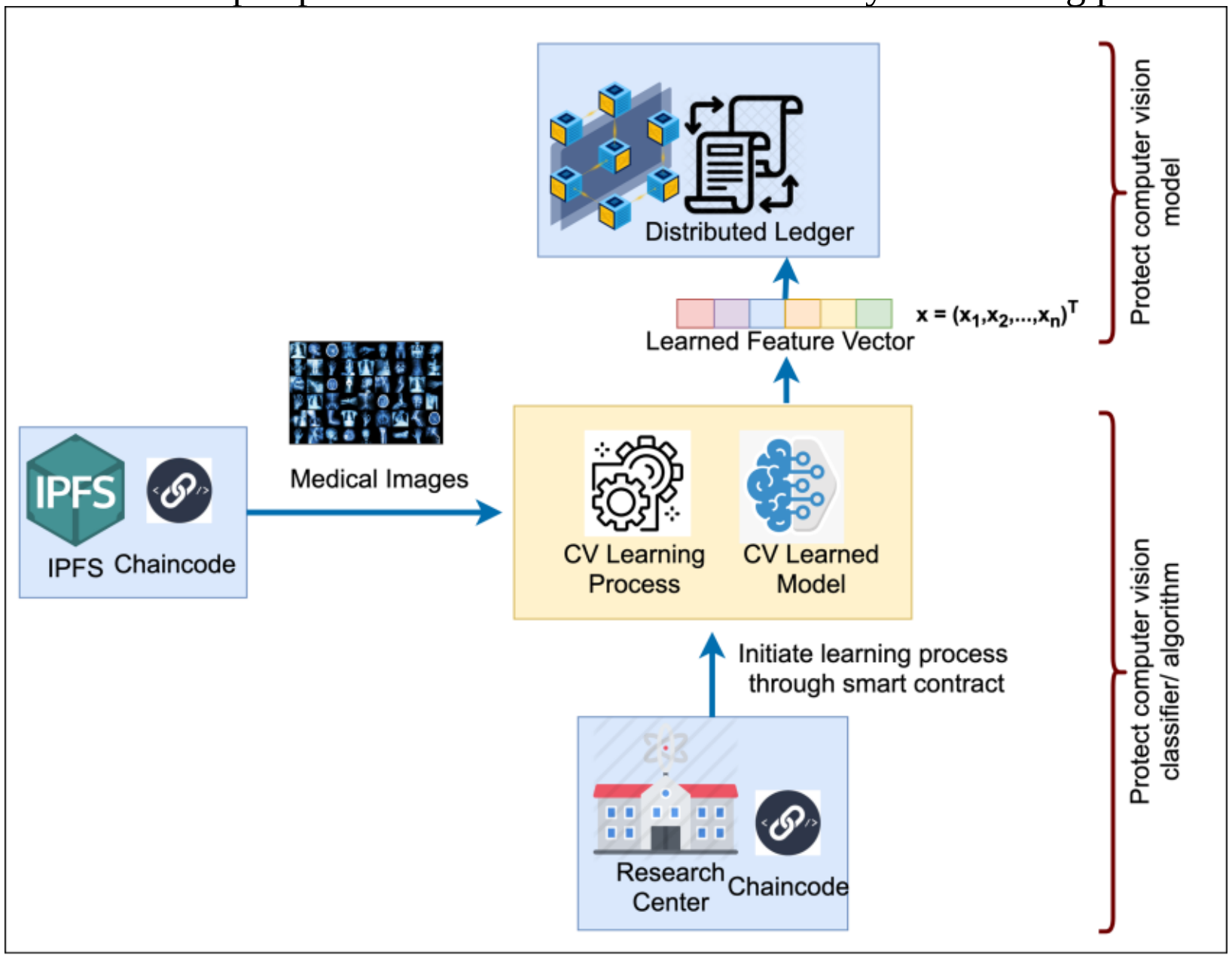

**Figure 21.** Blockchain Solution to protect training phase of Computer Vision.

3. Post Training Phase:

In the post-training phase of computer vision-based healthcare, the result generated for given image input should be self-explanatory, describing justification for output. Access to the trained model is restricted through smart contracts. Only authorized doctors, researchers can access the model. It will check with stored feature vector in distributed ledger for validating the integrity of model run. Here we are introducing explainable AI to make healthcare applications more trustworthy and transparent. Explainable AI (XAI) is a novel concept in machine learning that explains how AI systems make black box choices. XAI examines and attempts to clarify the stages and models making decisions. XAI prepares the description for output generated for the given input image. The metadata of those descriptions is stored in the blockchain for further verification and validation. This is how we will get a tamper-proof and accurate comprehensive diagnosis with explanation in a secure blockchain environment as illustrated in Figure 22.

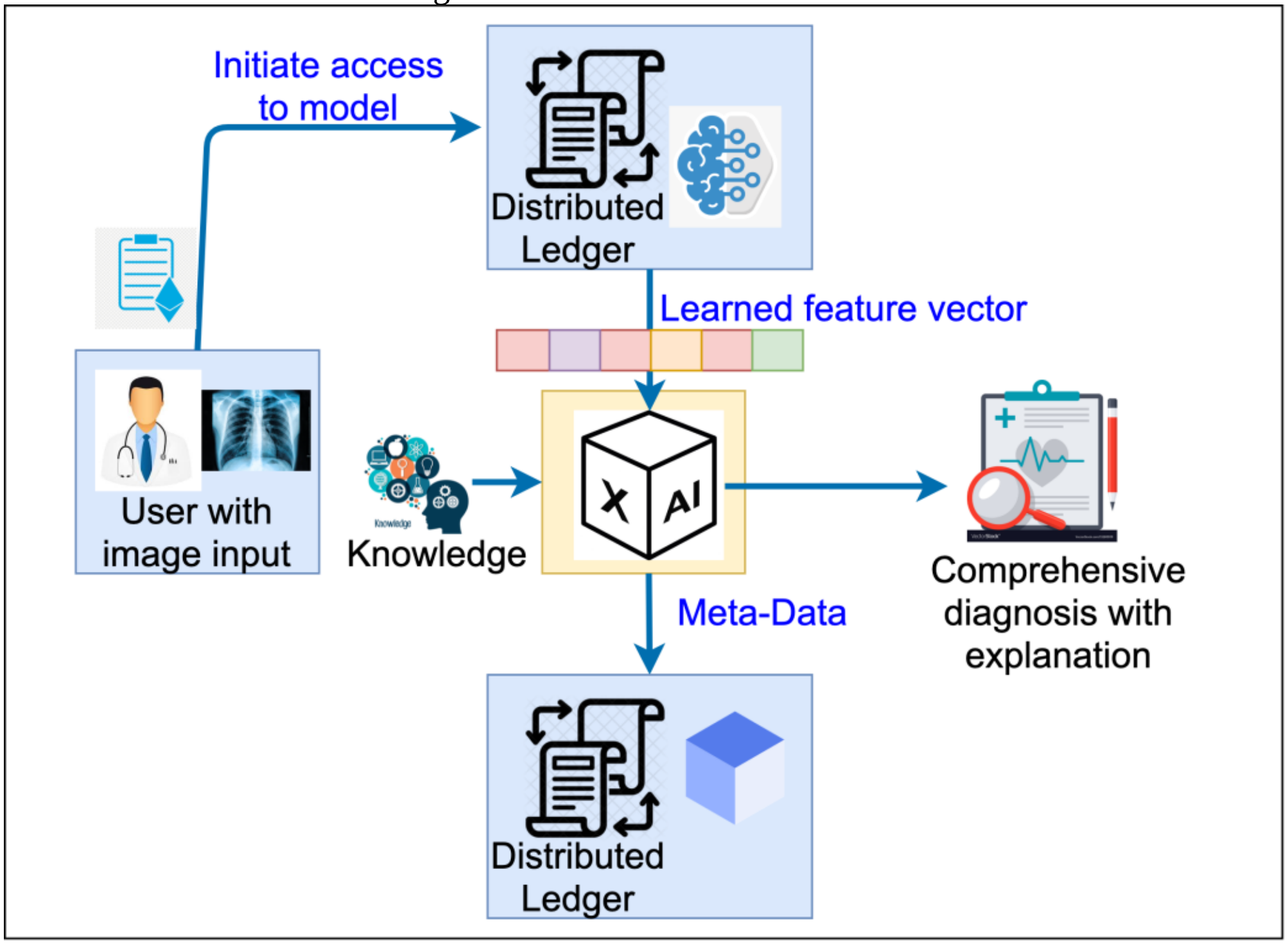

**Figure 22.** Post Training Blockchain Solution for Computer vision-based Healthcare.

## C. Acoustic AI-based Healthcare

Most recent advancements and transformational possibilities of machine learning (ML), particularly deep learning, in acoustics. ML is data-driven in comparison to traditional acoustics and signal processing. Thus, acoustic AI can have profound potential in analyzing acoustic signals in healthcare applications. However, as discussed in the above section, healthcare applications have the biggest threat of adversarial attacks, which are harmful in the healthcare domain. Adversarial attacks on acoustic signals are complex and have a moderate level of perceivability. Therefore, we can keep acoustic medical data on local machines and impose federated learning with blockchain technology to protect acoustic AI-based healthcare applications from adversarial attacks.

**Synthesized Framework:**

1. Dataset Building:

In acoustic AI, data can reside on local machines of data owners like doctors, hospitals, laboratories, etc., as we have framed blockchain solution for NLP-based healthcare. The following Figure 23 illustrates the dataset building with blockchain for acoustic AI-based healthcare. Various IoMT devices can threaten security at each data owner as rogue devices can enter the IoMT network, affecting the data generation process. Hence rogue device mitigation strategy based on blockchain can be introduced in dataset building. Through smart contracts, access control rules will be implied on acoustic data storage, and IoMT devices must go through proper registration and authentication procedure to contribute to dataset building. Data sharing for dataset building in acoustic AI-based healthcare is like the NLP-based healthcare dataset building framework described earlier.

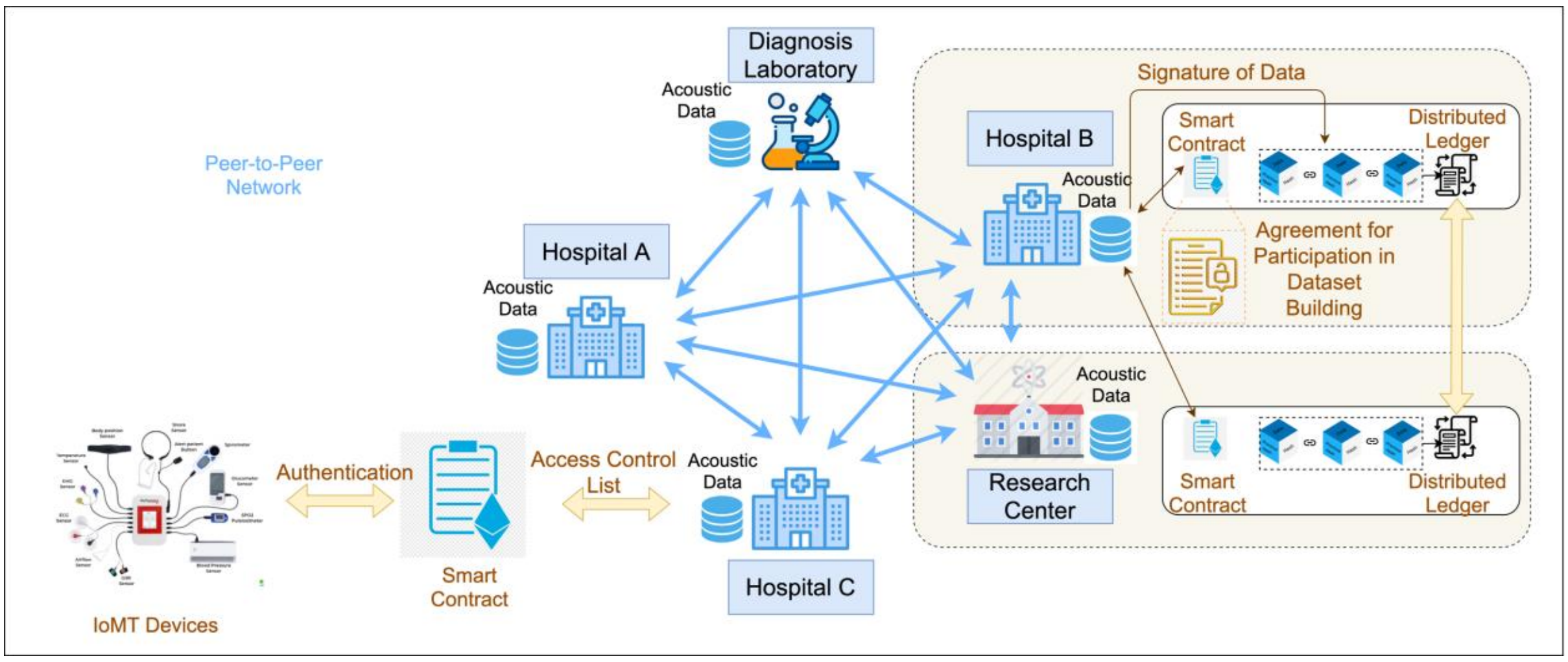

**Figure 13.** Dataset Building in Acoustic AI-based Healthcare with Blockchain.

2. Training Phase:

As we are keeping acoustic data in distributed form, a federated learning approach is adopted in our secure framework of training acoustic AI model. Each data owner acts as a node in the network running the local acoustic AI model. The extracted features from audio samples will be kept secure in distributed ledgers for further validation and reference. Extracted features of audio samples can be any form based on the learning approach. Suppose we are using a traditional machine learning approach. In that case, it can be Amplitude Envelope, Zero-Crossing Rate (ZCR), Root Mean Square (RMS) Energy, Spectral Centroid, Band Energy Ratio, and Spectral Bandwidth. For the deep learning approach, it can be spectrograms, Mel-spectrograms, and Mel-Frequency Cepstral Coefficients (MFCCs). Each local gradient is stored in blockchain, and consensus algorithms drive the global model generation task. The learned global model is then available in the distributed ledger, preventing unauthorized access to the model. This is how acoustic AI classifier/algorithm and model is protected with blockchain technology as described in Figure 24.

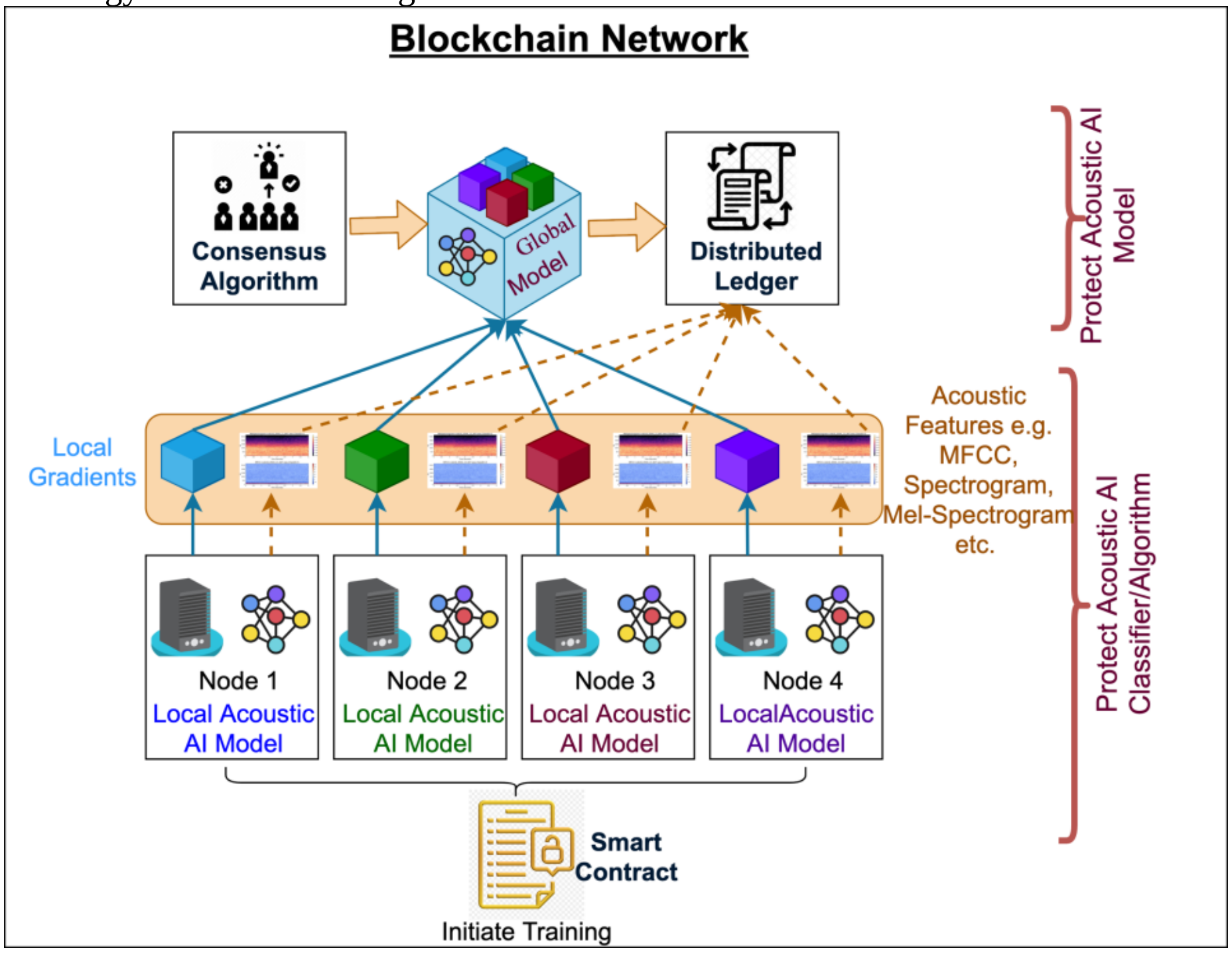

**Figure 14.** Protecting Training Phase of Acoustic AI-based Healthcare with Blockchain.

3. Post Training Phase:

In the post-training phase of acoustic AI-based healthcare application, using blockchain adversarial input to the model will be identified and keeping secure learned model by restricting access through smart contracts as shown in Figure 25. Temporal dependency is an important characteristic of audio signals. We are using these characteristics and a consensus algorithm to detect whether the input to the model is genuine or adversarial. The following Figure demonstrates the blockchain framework designed for protecting the post-training phase of acoustic AI-based healthcare. Firstly, a smart contract initiates access to the model and gets the result for audio input provided by the user. Then original input is passed to the consensus algorithm in blockchain network. It will split the input in N number of segments and assign each segment to each miner available in the miner pool. Miners will again execute the model for segments, and at the end, all miners consensually combine the result generated for each segment. Due to temporal dependency, if that input was adversarial, then aggregated results will not make any sense and will be different from the generated output at the first step. This strategy will help detect adversarial input to the model and increase trust in the model.

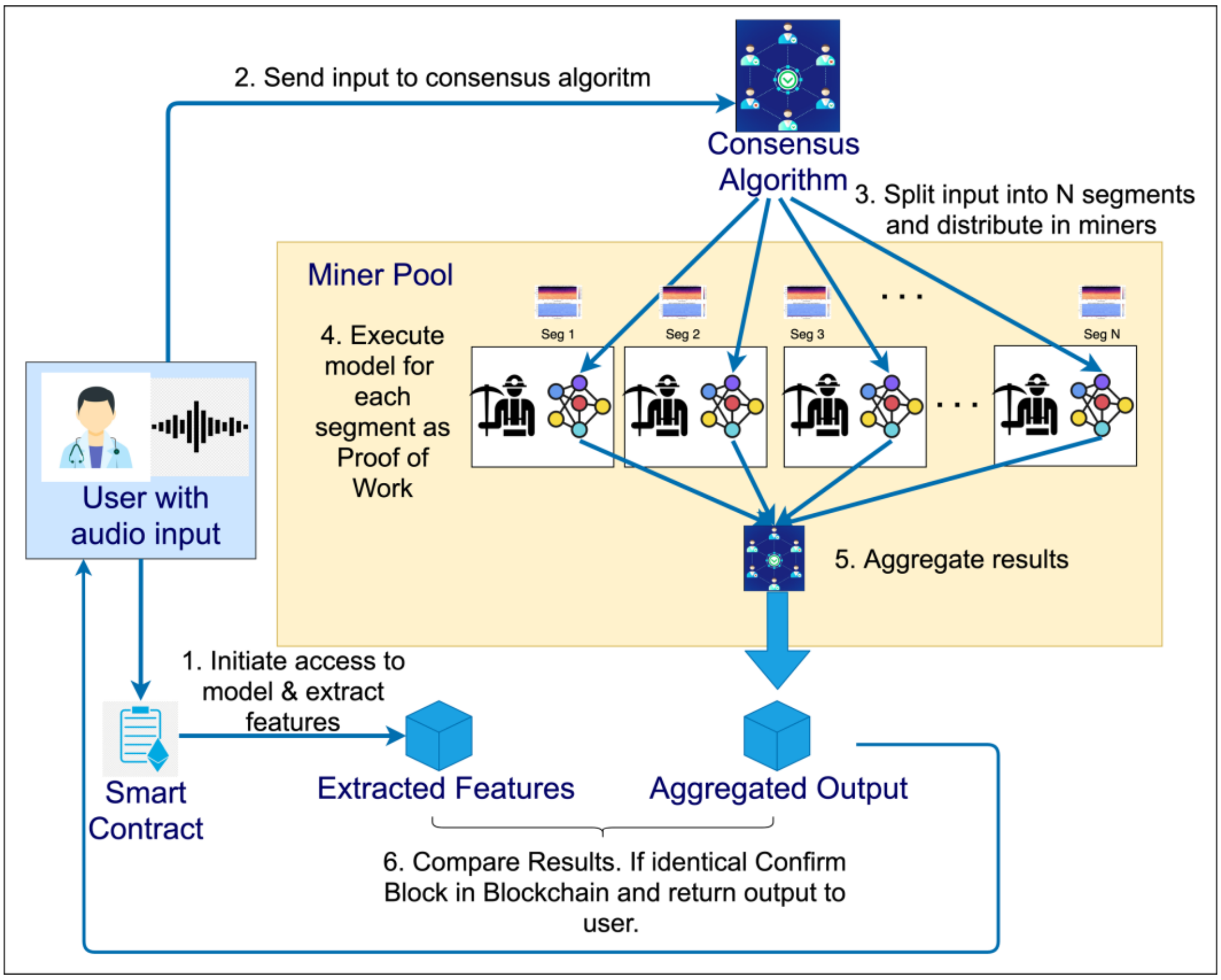

**Figure 15.** Protecting Post Training Phase of Acoustic AI-based Healthcare with Blockchain.

## X. Discussion

Blockchain technology gathers real-time data, stores it on several servers to keep it hack resistant, limits access to only approved individuals, and maintains an updated version of the file every time it is visited.

### A. The Survey Outcome:

This survey aids in comprehending the significance of blockchain in AI-based healthcare applications. According to a literature review, blockchain technology offers more promising security, privacy, and trust in AI-based healthcare applications. To help improve the wide acceptance and making more robust AI applications in healthcare, authors have synthesized a blockchain framework that can mitigate adversarial attacks considering the need for individual Natural Language Processing, Computer Vision, and Acoustic AI domains. Furthermore, authors have applied knowledge of blockchain features and properties to protect datasets, classifier/algorithm, and AI model from adversarial attacks. Following Figure 26 provides detailed applicability of blockchain properties from protecting and validating datasets, protecting classifiers/algorithms protecting the post-training environment in AI.

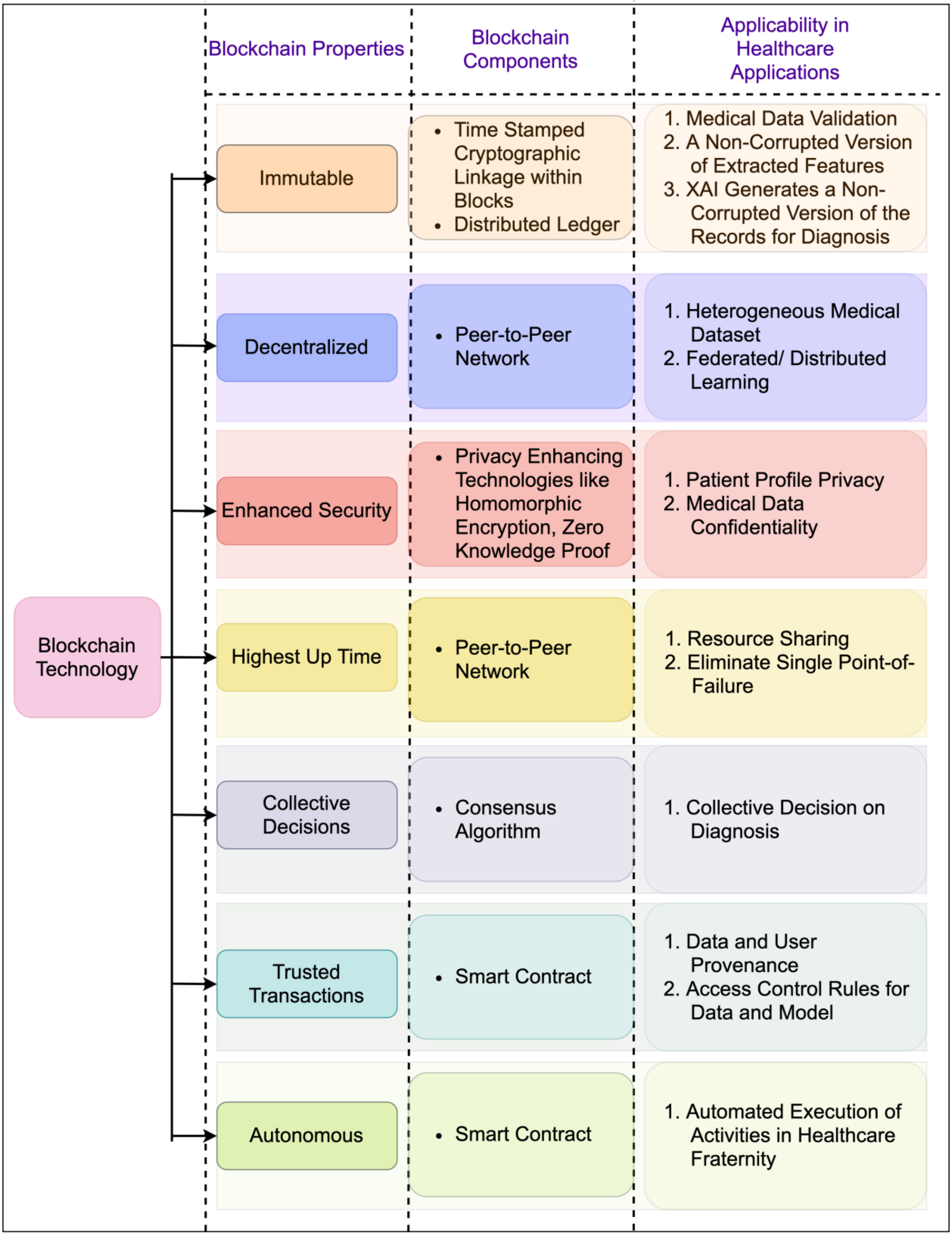

**Figure 16.** Blockchain for AI-based Healthcare explained.

**B. Challenges in the Healthcare Industry to Adopt Blockchain Technology in India**

Due to healthcare's highly regulated environment and low-risk appetite, the healthcare industry often adopts digital technology late.

1. Lack of Computational Environment

Blockchain requires much computational power to keep transaction information encrypted. The blockchain solution uses much electricity and generates heat. India has an electric supply shortage, and the country's typical temperature is warm. India may consider decreasing the number of servers or employing a single server to conserve power. Blockchain storage facilities should be built in cooler areas with internet access. Most healthcare facilities lack computers, patient data are handwritten, and internet access is inconsistent. The expense of operating such a system must be shared equally among the government, healthcare providers, pharmaceutical companies, insurance companies, and other stakeholders.

2. Lack of Uniform IT System and Interoperability

There is currently no unified IT system for healthcare in India. Interoperability difficulties arise because of several digital solutions. Interoperability is particularly difficult to establish since various healthcare facilities employ different models and have their own set of codes. When patients switch service providers for unanticipated reasons, the problem becomes much worse. Patients are compelled to repeat diagnostic tests and treatment procedures, increasing overhead expenses and patient dissatisfaction.

## C. Advancements in Blockchain

1. Quantum Blockchain

The term "quantum blockchain" can be considered a decentralized, encrypted, and distributed database reliant on quantum computing and quantum information theory. The key features of quantum blockchain are security and efficiency. Quantum secure direct communication (QSDC) or quantum key distribution (QKD) can be used to ensure communication security between nodes. As a result, the characteristics of quantum physics provide network authentication. Traditional encryption techniques used in digital signatures, such as RSA, might be deceptive in the face of quantum computer attacks. The quantum digital signature technique can be employed in quantum blockchain to overcome this problem. As a result, the quantum blockchain possesses quantum security properties. As a result, quantum computers might not be able to attack the quantum blockchain. The blockchain with quantum technology also can execute transactions quickly [127]–[129].

2. Hyperledger

Hyperledger is specifically a blockchain-based platform that is being utilized to incorporate healthcare data management developments. Healthcare applications built on Hyperledger Fabric have emerged as a popular blockchain deployment. The linux foundation hosts an open-source community called Hyperledger, seeking to offer suitable foundations, rules, libraries, and tools for creating global business blockchain projects. Hyperledger is ideally suited for healthcare applications. It also provides complete control over smart contracts written in several computer languages, including Node.js and JavaScript. Bitcoin and Ethereum, on the other hand, can execute seven and fifteen transactions per second, respectively. With transaction speeds of up to 3000 transactions per second, Hyperledger beats this competition. Another benefit is that it has a high transaction throughput and low transaction cost. Compared to other blockchain frameworks, Hyperledger Fabric is the most sophisticated blockchain framework available[130].

3. Zero-Knowledge Proof (ZKP)

Zero-Knowledge Proof (ZKP) has been used as one of the most effective methods for meeting transaction secrecy requirements. The purpose of ZKP's functionality is to allow a prover to persuade a verifier of a certain truth without exposing the real information. Interactive zero-knowledge proof and non-interactive zero-knowledge proof are the two most significant forms of ZKPs. Interactive ZKPs require the prover to execute a sequence of activities to convince the verifier of a certain truth. The sequence of activities in interactive ZKPs is linked to mathematical probability principles. Non-interactive ZKPs, on the other hand, do not include any interactive elements. As in non-interactive ZKPs, the prover may generate all challenges simultaneously, and the verifier could reply afterwards. Zero-knowledge- Succinct Non-Interactive Argument of Knowledge (zk-SNARKs), are another significant type of ZKPs that has recently developed. Among the zero-knowledge proof alternatives, zk-SNARK is the most popular [131]. It can assist in the definition of a quadratic equation that uses public, private, and input data to generate evidence.

4. Homomorphic Encryption

Among the most secure privacy enhancing techniques for conducting operations on encrypted data is homomorphic encryption. The operations on encrypted data would provide the same outcomes as if they were done on unencrypted data. As a result, organizations may employ homomorphic encryption for data analysis without sacrificing data anonymity or privacy. For privacy-preserving outsourced storage and computing, homomorphic encryption can be leveraged. Homomorphic encryption makes it possible to encrypt data before sending it to commercial cloud environments for processing. Homomorphic encryption may be thought of as an evolution of asymmetric-key or public-key cryptography. The encryption and decryption functions may be thought of as homomorphisms between plaintext and ciphertext spaces, like homomorphisms in algebra [132], [133]

5. Other Distributed Ledger Technologies

Systemic inadequacies and scalability issues drove researchers to seek solutions outside of the blockchain. As a result, there have been innovative and creative inventions such as hashgraph, directed acyclic graph (DAG), and holochain [134], [135]. In general, the goal is to maintain blockchain's original purpose alive in the context of different challenges.

*a) Hashgraph*

Hashgraph is a form of distributed ledger technology based on consensus building. The distributed ledger technology (DLT) especially depends on consensus timestamping to ensure that network transactions agree with every node in the network. The consensus algorithm highlights the network's stability and excellence in distributed ledger technology. This sort of DLT network, unlike typically distributed ledger technology networks, achieves transaction success exclusively by consensus. Nodes on the hashgraph DLT can encounter fairness thanks to virtual voting and gossip about gossip methods. Consensus timestamping prevents blockchain issues such as transaction cancellation or inclusion on future blocks. Since there is no requirement for proof of work on this DLT network, there may be thousands of TPS.

*b) DAG*

DAG is a distributed ledger technology that uses consensus methods. Consensus algorithms function in such a fashion that transactions that succeed just require the network's majority support. There is considerably more collaboration and teamwork in such a network, and nodes possess equal opportunities. The primary goal of a DLT was to democratize the internet economy. A private blockchain network, for example, is led by a centralized authority, which removes democracy from the DLT. On the other hand, this distributed ledger technology offers equal weight to every node in the network. As a result, each node is not required to refer to each other. IOTA's Tangle is one of the most famous "current generation" networks that use the DAG data structure. In this case, miners/nodes can undertake dual functions that nodes in the blockchain do independently. This means that a Tangle miner can originate and approve a transaction at the same time.

*c) Holochain*

Holochain aspires to build a distributed network that will serve as the foundation for the “next-generation internet.” Holochain is a hybrid of blockchain, BitTorrent, and Github. This is a DLT that distributes data across nodes to limit centralized control over data flow. This distributed platform essentially means that each node will run on its chain. This means that nodes or miners are free to operate individually. In addition, users can store data using specific keys in what the holochain team refers to as a distributed hash table (DHT). This data, however, remains in physical places "distributed” throughout the world.
Following Table 13 provides a comparison within these Distributed Ledger Technologies.

**Table 11.** Difference Distributed Ledger Technologies.

| Parameter | Blockchain | Hashgraph | DAG | Holochain |
|---|---|---|---|---|
| Launched in | 2008 | 2018 | 2015 | 2018 |
| Consensus Algorithm | Many Algorithms Available | Virtual Voting | The previous Transaction validates the new one | Not Required |
| Scalability | Limited | High | High | Infinite |
| Transaction Execution Speed | Limited | High | High | Highest |
| Data Structure | Blocks are generated as per the sequence of transactions. | Gossip about Gossip Protocol | Directed Acyclic Graph | Distributed among Nodes |
| Examples | Bitcoin, Ethereum, etc. | Swirlds & NOIA | NXT, Tangle, ByteBall, etc. | Holochain |

# XI. Conclusion

The objective of this SLR is to review the work done in blockchain for robust AI-based healthcare. It has been found that very little literature is available in this area. In this study, authors have also presented literature from various categories as Natural Language Processing, Computer Vision, and Acoustic AI in healthcare. Further, we reviewed articles on the healthcare applications and adversarial attacks they might go through for each category. The potential for blockchain to change the healthcare business is undeniable. Without sacrificing security, blockchain may be utilized to protect patients' access to complete medical treatment and as a fantastic method to keep the healthcare system's whole supply chain functioning smoothly. The numerous characteristics of the blockchain technology point to one undeniable reality of the potential to balance the conflict between data sharing and privacy, especially in AI-based healthcare. Future research directions have been presented in a synthesized blockchain framework for NLP, Computer Vision, and Acoustic AI, which was created by combining insights from existing applications, adversarial attacks, and threats in existing technology outlined throughout this review for AI-based healthcare.